\documentclass[trackchanges,twocolumn]{aastex701}
\usepackage{xcolor}

\newcommand{\source}{1E 1547.0$-$5408}
\newcommand{\ixpe}{\emph{IXPE}}

\renewcommand\thetable{\arabic{table}}

\begin{document}

\title{The long quest for vacuum birefringence in magnetars: \source\ and the elusive smoking gun}

\author[0000-0003-3259-7801]{Roberto Taverna}
\affiliation{Department of Physics and Astronomy, University of Padova, via Marzolo 8, I-35131 Padova, Italy}
\email{taverna@pd.infn.it}

\author[0000-0003-3259-7801]{Roberto Turolla}
\affiliation{Department of Physics and Astronomy, University of Padova, via Marzolo 8, I-35131 Padova, Italy}
\affiliation{Mullard Space Science Laboratory, University College London, Holmbury St Mary, Dorking, Surrey RH5 6NT, UK}
\email{turolla@pd.infn.it}

\author[0009-0001-4644-194X]{Lorenzo Marra}
\affiliation{INAF - Istituto di Astrofisica e Planetologia Spaziali, Via del Fosso del Cavaliere 100, I-00133, Roma}
\email{lorenzo.marra@inaf.it}

\author[0000-0002-5004-3573]{Ruth M.E. Kelly}
\affiliation{Mullard Space Science Laboratory, University College London, Holmbury St Mary, Dorking, Surrey RH5 6NT, UK}
\email{rmek@mssl.ucl.ac.uk}

\author[0000-0001-8785-5922]{Alice Borghese}
\affiliation{European Space Agency (ESA), European Space Astronomy Centre (ESAC), Camino Bajo del Castillo s/n, 28692 Villanueva de la Ca\~{n}ada, Madrid, Spain}
\email{alice.borghese@gmail.com}

\author[0000-0001-5480-6438]{Gian Luca Israel}
\affiliation{INAF - Osservatorio Astronomico di Roma, via Frascati 33, I-00078 Monteporzio Catone, Italy}
\email{gianluca.israel@inaf.it}

\author[0000-0003-3259-7801]{Sandro Mereghetti}
\affiliation{INAF - Istituto di Astrofisica Spaziale e Fisica Cosmica di Milano, via Corti 12, I-20133 Milano, Italy}
\email{sandro.mereghetti@inaf.it}

\author[0000-0001-5902-3731]{Andrea Possenti}
\affiliation{INAF - Osservatorio Astronomico di Cagliari, via della Scienza 5, I-00947 Selargius, Italy}
\email{andrea.possenti@inaf.it}

\author[0000-0001-5326-880X]{Silvia Zane}
\affiliation{Mullard Space Science Laboratory, University College London, Holmbury St Mary, Dorking, Surrey RH5 6NT, UK}
\email{sz@mssl.ucl.ac.uk}

\author[0000-0001-6641-5450]{Michela Rigoselli} 
\affiliation{INAF - Osservatorio Astronomico di Brera, via Brera 28, I-20121 Milano, Italy}
\affiliation{INAF - Istituto di Astrofisica Spaziale e Fisica Cosmica di Milano, via Corti 12, I-20133 Milano, Italy}
\email{michela.rigoselli@inaf.it}

\begin{abstract}

Magnetars are now known to be among the most strongly polarized celestial sources in X-rays. Here we report on the $500\,\mathrm{ks}$ observation of the magnetar \source\ performed by the Imaging X-ray Polarimetry Explorer (\ixpe) in March 2025. The \ixpe\ spectrum is well reproduced by a single thermal component with  blackbody temperature $kT_\mathrm{BB}\sim 0.67\,\mathrm{keV}$ and emission radius $R_\mathrm{BB}\sim 1.2\,\mathrm{km}$. The source exhibits a high linear polarization degree in the $2$--$6\,\mathrm{keV}$ band ($\mathrm{PD}=47.7\pm2.9\%$) with polarization angle $\mathrm{PA}=75^\circ.8 \pm 1^\circ.8$, measured West of celestial North. While $\mathrm{PA}$ does not appear to vary with energy, there is some evidence (at the $1\sigma$ confidence level) of a minimum in $\mathrm{PD}$ between $3$ and $4\,\mathrm{keV}$, compatible with what is expected by partial mode conversion at the vacuum resonance in a magnetized atmosphere. Phase-resolved spectral and polarimetric analyses reveal that X-ray thermal radiation likely originates from a single, fairly small hot spot with a non-uniform temperature distribution. Fitting the phase-dependent $\mathrm{PA}$ measured by \ixpe\ with a rotating vector model (RVM) constrains the source geometry and indicates that both the dipole axis and line-of-sight are misaligned with respect to the spin axis. Under these conditions, the high polarization of the source cannot be regarded as compelling evidence for the presence of vacuum birefringence in the star magnetosphere. Nevertheless, the fact that the RVM successfully reproduces the modulation of the X-ray polarization angle and the behavior of  $\mathrm{PD}$ with the energy hint once more to the presence of QED effects in magnetars.

\end{abstract}


\section{Introduction}\label{sec:intro}
\setcounter{footnote}{0}

Soft gamma repeaters (SGRs) and anomalous X-ray pulsars (AXPs) form a distinct class of Galactic X-ray pulsars characterized by long rotational periods, 
$P\approx 1$--$12\,\mathrm s$, and large spin-down rates, 
$\dot{P} \approx 10^{-15}$--$10^{-10}\,\mathrm{s\,s^{-1}}$. These sources produce recurrent, sub-second bursts of hard X-ray/soft-$\gamma$ radiation and exhibit persistent X-ray luminosities of $L_{\mathrm{X}} \approx 10^{30}$--$10^{35}\,\mathrm{erg\,s^{-1}}$, typically exceeding their rotational energy losses. 
Their timing properties imply surface dipole magnetic fields of 
$B \approx 10^{13}$--$10^{15}\,\mathrm{G}$, and the absence of detectable companions identifies them as magnetars: neutron stars powered primarily by the decay and dissipation of ultra-strong magnetic fields \cite[][see also \citealt{turolla+15,kaspi+belo17} for reviews]{td92,Td93}.

Below $\sim 10\,\mathrm{keV}$, magnetar spectra are commonly represented by two-component models, either two blackbodies or a blackbody plus a power law. The thermal components arise from localized hot spots on the neutron star surface, whereas the non-thermal component is produced via resonant Compton scattering (RCS) of surface photons by magnetospheric charges in a twisted magnetic field \citep{tlk02,fernandez2007,ntz08}. Many magnetars also emit a hard X-ray tail extending up to $\sim 100$--$200\,\mathrm{keV}$, often with strong pulsations.

In such extreme magnetic environments, the propagation of electromagnetic radiation is profoundly modified, as first discussed in the pioneering work of \citet[see also \citealt{1997JPhA...30.6485H} and \citealt{2006RPPh...69.2631H} for a comprehensive review]{1978JETPL..27..305G,1978SvAL....4..117G}. The combined plasma–vacuum dielectric tensor becomes highly anisotropic, and the ordinary (O) and extraordinary (X) polarization eigenmodes acquire markedly different opacities. Radiative-transfer calculations predict correspondingly large linear polarization fractions, up to
$\sim 80\%$ and $\sim 20\%$, from thermal emission emerging from strongly magnetized atmospheres or condensed surfaces, respectively \citep[e.g.,][]{Fernandez2011,Taverna+14,Taverna+20}. Until recently, these predictions remained untested, owing to the lack of sensitivity in X-ray polarimeters.

The launch of the Imaging X-ray Polarimetry Explorer (\ixpe) in December 2021 \citep{Weisskopf2022} enabled the first high-quality spectro-polarimetric measurements in the $2$--$8\, \mathrm{keV}$ band. During its four years of operations, \ixpe\ observed about one hundred X-ray sources, including five persistent magnetars, \object{4U 0142$+$61} \citep{taverna+22}, \object{1RXS J170849.0$-$400910} \cite[hereafter 1RXS J1708;][]{zane+23}, SGR 1806$-$20 \citep{turolla+23}, 1E 2259$+$586 \citep{heyl+24} and 1E 1841$-$045 \citep{rigoselli+25, stewart+25a}. With the exception  of SGR 1806$-$20, statistically significant polarization (energy- and phase-integrated) was detected, ranging from $\approx 6\%$ to $\approx 35\%$. 
The degree of polarization is strongly dependent on the energy: it is moderate ($\approx 10$--$20\%$) at $2$--$3\, \mathrm{keV}$ 
and then increases to $\approx 60$--$80\%$ at $6$--$8\, \mathrm{keV}$ in 1E 1841$-$045 and 1RXS J1708, or to $\approx 35\%$ in 4U 0142$+$61. 
The polarization angle remains nearly constant with the energy in the former two sources, implying that the same polarization mode dominates through the \ixpe\ band. In contrast, 4U 0142$+$61 exhibits a drop in the polarization degree to zero near $4\, \mathrm{keV}$ accompanied by $\sim 90^\circ$ rotation of the polarization angle, signaling a transition in the prevailing polarization mode \citep[see][for an extensive discussion and \citealt{lai23} for an alternative interpretation]{tatu24}. 

In this paper, we report on the spectral and polarization analysis of the \ixpe\ observation of the magnetar \source, targeted in the GO2 observing run (P.I. George Younes). We provide a comparative assessment of our findings against those recently presented in a preprint by \citet{2025arXiv250919446S} and highlight similarities and differences between their (preliminary) analysis and the present one.
Observations and data processing are detailed in Section \ref{sec:obs-ixpe}. The results of our spectral, timing, and polarimetric analyzes are presented in Section \ref{sec:res} and discussed in Section \ref{sec:discussion}. Conclusions follow in Section \ref{sec:concl}.

\section{Observations and data analysis} \label{sec:obs-ixpe}


The \ixpe\ observation of \source\ started on 2025 March 26 03:01:52 UTC and ended on 2025 April 5 00:26:53 UTC, for a total on-source exposure time of $\approx500\,\mathrm{ks}$. After retrieving the level 1 and level 2 data files from the HEASARC \ixpe\ public archive\footnote{\url{https://heasarc.gsfc.nasa.gov/docs/ixpe/archive}}, we referred the event arrival times to the Solar System barycenter using the ftool \texttt{barycorr} with JPL planetary ephemerids DE430 \cite[$\mathrm{RA}=15^\mathrm{h}\,50^\mathrm{m}\,54.12386^\mathrm{s}$, $\mathrm{DEC}=-54^\circ\,18^\prime\,24.1131^{\prime\prime}$, taken from the ATNF Pulsar Catalogue\footnote{\url{https://www.atnf.csiro.au/research/pulsar/psrcat}},][]{2005AJ....129.1993M}, and performed proper background rejection, according to the procedure discussed in \citet{2023AJ....165..143D}; furthermore, some time intervals containing the most prominent background spikes, for a total of $\lesssim100\,\mathrm{s}$, were removed from the data set. We then extracted from the resulting photon lists the source counts from a circle of radius $r^\mathrm{src}$, centered on the position of the source and subtracted the background counts taken from a concentric annular region, with inner and outer radii $r^\mathrm{bkg}_\mathrm{inn}$ and $r^\mathrm{bkg}_\mathrm{out}$, respectively. Following the procedure outlined in \citet{2004MNRAS.351..161P}, we find that the best signal-to-noise ratio is obtained for $r^\mathrm{src}=60^{\prime\prime}$, $r^\mathrm{bkg}_\mathrm{inn}=120^{\prime\prime}$ and $r^\mathrm{bkg}_\mathrm{out}=240^{\prime\prime}$; this produced about $25,600$ source counts summed over the three \ixpe\ detector units (DUs). By taking $r^\mathrm{src}\lesssim 50^{\prime\prime}$ an increasing fraction of the source photons is lost, up to $\sim 20\%$ when $r^\mathrm{src}$ shrinks to $\sim30^{\prime\prime}$.

The source counts are completely dominated by the background above $\approx\!6\,\mathrm{keV}$,
preventing a meaningful spectro-polarimetric analysis at higher energies (see \S\ref{sec:spectral-pa}). We therefore opted to restrict our investigation to the $2$--$6\,\mathrm{keV}$ energy band and to use the weighted analysis procedure. To this end, we processed the level 2 files using the latest version of the \ixpe\ response functions (20250225), available online in the HEASARC calibration database\footnote{\url{https://heasarc.gsfc.nasa.gov/docs/ixpe/caldb}}.

\section{Results}\label{sec:res}
\subsection{Timing analysis} \label{sec:timing}

We processed the level 2 photon lists using version 8.1 of the \textsc{hendrics} package \cite[][see also \citealt{2015ApJ...800..109B} for a complete description of the \textsc{stingray} software]{2018ascl.soft05019B}, to compute the timing solution for \source. The most prominent peak in the frequency spectrum appears in the $0.45$--$0.50\,\mathrm{Hz}$ interval, within which we then performed a $Z^2_n$ search. Sinusoidal pulsations were detected at a frequency $\nu_0=0.47785498(2)\,\mathrm{Hz}$, with a frequency derivative $\dot{\nu}_0=-4.7(6)\times10^{-12}\,\mathrm{Hz/s}$. The times-of-arrival (TOAs) of the events were then calculated using the \texttt{HENphaseogram} tool, dividing the observation into $75$ time intervals. Finally, by fitting the TOAs  with a linear spin-down relation $\nu=\nu_0+\dot{\nu}_0t$ with the \textsc{pint} software \cite[v1.1.4,][]{2021ApJ...911...45L}, we obtained the best-fitting spin frequency and frequency derivative as $\nu=0.477855133\pm2.1\times10^{-8}\,\mathrm{Hz}$ and $\dot{\nu}=-(4.68\pm0.19)\times10^{-12}\,\mathrm{Hz/s}$ ($\chi^2/\mathrm{dof}=40.832/76$), respectively, at epoch $60765.0724\,\mathrm{MJD}$; 
the corresponding values of the period and period derivative are $P=2.092684438\pm9.4\times10^{-8}\,\mathrm{s}$ and $\dot{P}=(2.05\pm0.08)\times10^{-11}\,\mathrm{s/s}\,$\footnote{Here and in the following, errors are quoted at $1\sigma$ confidence level, unless explicitly stated otherwise.} (for a spin-down dipole magnetic field strength $B_\mathrm{sd}\approx2\times10^{14}\,\mathrm{G}$ at the magnetic equator). This timing solution was then used  to phase-fold the photon lists with the \texttt{xpphase} tool within the \textsc{ixpeobssim} package \cite[][see Section \ref{sec:phaseres}]{2022ascl.soft10020B}.

\subsection{Phase-averaged spectral analysis} \label{sec:spectral-pa}

\begin{figure*}[]
\includegraphics[width=18cm]{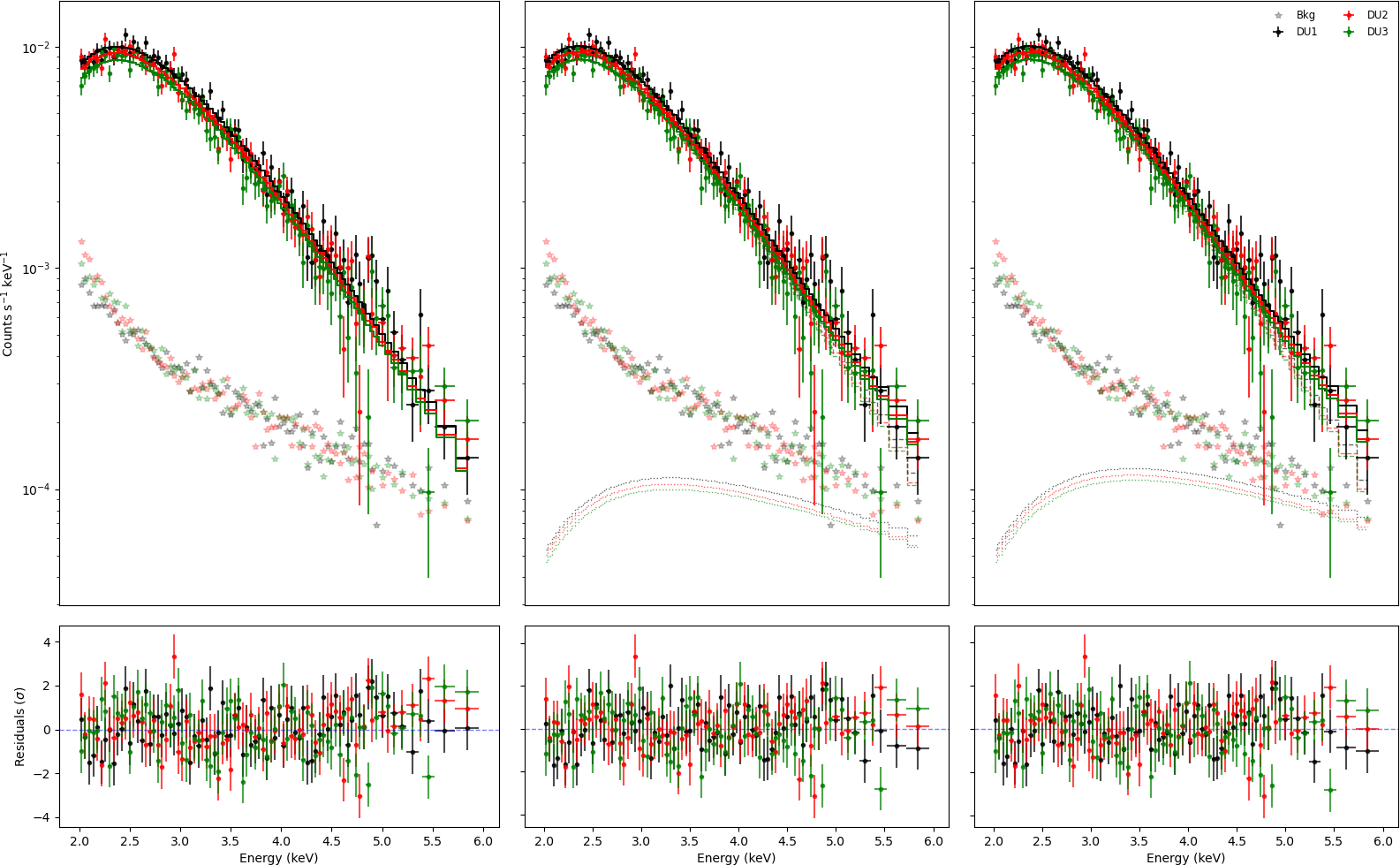}
\caption{\ixpe\ source (dots with error bars) and background (stars) spectra (top panel) of \source\ collected by DU1 (black), DU2 (red) and DU3 (green), and fitting residuals in units of the standard deviation (bottom panels) for the \texttt{tbabs}$\times$\texttt{bbodyrad} model with $N_\mathrm{H}$ frozen at $4.6\times10^{22}\,\mathrm{cm}^{-2}$ (left), \texttt{tbabs}$\times$\texttt{(bbodyrad}$+$\texttt{bbodyrad)} model with $N_\mathrm{H}$ frozen at $4.6\times10^{22}\,\mathrm{cm}^{-2}$ (center) and \texttt{tbabs}$\times$\texttt{(bbodyrad}$+$\texttt{powerlaw)} model with $N_\mathrm{H}$ frozen at $4.9\times10^{22}\,\mathrm{cm}^{-2}$ (right). The best fitting models (solid histograms) and the individual components (dotted histograms) are also shown in each plot. Fit results are summarized in Table \ref{tab:spec_xspec}.
\label{fig:pa_spectra}}
\end{figure*}

We processed data from the three \ixpe\ DUs within \textsc{xspec} \cite[][]{1996ASPC..101...17A}, generating the Ancillary Response Files (ARFs) and Modulation Response Files (MRFs) with the \textsc{ixpecalcarf} tool \cite[][]{2025AAS...24526001C}. Data were grouped to ensure a signal-to-noise ratio $\gtrsim3$ in each bin
in the $2$--$6\,\mathrm{keV}$ energy band. A normalization factor was included in all the models to account for the different cross-calibration of the three DUs (assuming unity for DU1). Interstellar absorption was modeled using \texttt{tbabs}, with abundances taken from \cite{2000ApJ...542..914W} and photoionization cross-sections from \citet{1996ApJ...465..487V}. We adopted the column density $N_\mathrm{H}=4.6\times10^{22}\,\mathrm{cm}^{-2}$ quoted in the systematic study by \citet{2020A&A...633A..31C}, based on \textit{XMM-Newton}, \textit{Chandra}, and \textit{Swift-XRT} data collected between 2009 and 2017 \citep[see][]{2018MNRAS.474..961C}, since the limited coverage of \ixpe\ at low energies does not allow a precise determination of $N_\mathrm{H}$.
We first attempted a fit using a single absorbed blackbody (BB) component (\texttt{tbabs}$\times$\texttt{bbodyrad}). The model shows a good agreement with the data (see Figure \ref{fig:pa_spectra} and Table \ref{tab:spec_xspec}), with $\chi^2/\mathrm{dof}=263.95/248$. 
The resulting blackbody temperature, $\approx0.67\,\mathrm{keV}$, is fully consistent with the values reported from observations taken shortly after the 2009 outburst of \source\ \cite[see][]{2011A&A...529A..19B}. However, the corresponding emission radius, $\approx 1.2\,\mathrm{km}$ for a distance of $4.5\,\mathrm{kpc}$ \cite[][]{2010ApJ...710..227T}, is somewhat smaller and more in line with the most recent estimates \cite[][]{2020A&A...633A..31C,2023ApJ...945..153L}. This discrepancy 
is most likely due to the larger size of the emitting region during an outburst phase, which then shrinks as the source returns in quiescence \citep[see e.g.][]{2022ApJ...936...99D}.

For completeness, and given that previous studies of the source at soft X-ray energies showed the presence of a second spectral component \cite[][]{2011A&A...529A..19B,2020A&A...633A..31C}, we also considered adding to the model either a blackbody or a power-law (PL). Keeping $N_\mathrm{H}$ fixed at the above value, the fit with two blackbodies, \texttt{tbabs}$\times$\texttt{(bbodyrad}$+$\texttt{bbodyrad)}, is also good, with $\chi^2/\mathrm{dof}=255.50/242$. 
The spectral parameters of the softer component are well constrained and similar to those obtained previously ($kT_1=0.65\,\mathrm{keV}$, $R_{{\mathrm {BB}}_1}=1.3\,\mathrm{km}$), while those of the hotter one are totally unconstrained. Similar results were obtained with a blackbody plus power-law model, \texttt{tbabs}$\times$\texttt{(bbodyrad}$+$\texttt{powerlaw)}, fixing $N_\mathrm{H}$ to the value obtained by \citet{2018MNRAS.474..961C} using a BB+2PL decomposition ($4.9\times10^{22}\,\mathrm{cm}^{-2}$). The fit is also statistically satisfactory ($\chi^2/\mathrm{dof}=253.94/242$), 
with blackbody temperature ($0.64\,\mathrm{keV}$) and radius ($1.4\,\mathrm{km}$) compatible with those obtained before, but the non-thermal component remains completely unconstrained (see again Table \ref{tab:spec_xspec} and Figure \ref{fig:pa_spectra} for details).  

We finally note that previous broad-band investigations of \source\   required the presence of an additional power-law component at hard X-ray energies, dominating the spectrum above $\approx10\,\mathrm{keV}$ \cite[][]{2011A&A...529A..19B,2020A&A...633A..31C,2023ApJ...945..153L}. We do not include this component in our fits, as its contribution in the \ixpe\ energy range is negligible. For all these reasons, we adopted the single-BB model in the remainder of our spectro-polarimetric analysis. Although our conclusions about the spectral model agree with those of \citet[][]{2025arXiv250919446S}, the values of $kT_\mathrm{BB}$ and $R_\mathrm{BB}$ are not compatible within $3\sigma$, possibly due to the different values assumed for $N_\mathrm{H}$.

\begin{table*}[ht!]
\tabletypesize{\scriptsize}
\begin{center}
\caption{Phase-averaged spectral analysis of the \ixpe\ data. \label{tab:spec_xspec}}
\begingroup
\setlength{\tabcolsep}{17pt}
\renewcommand{\arraystretch}{1.6}
\begin{tabular}{c | c c c}
\hline\hline
\  & \texttt{tbabs}$\times$\texttt{bbodyrad} & \texttt{tbabs}$\times$\texttt{(bbodyrad}$+$\texttt{bbodyrad)} & \texttt{tbabs}$\times$\texttt{(bbodyrad}$+$\texttt{powerlaw)} \\
\hline
$N_\mathrm{H}$ ($10^{22}\,\mathrm{cm}^{-2}$) & $4.6$ $^\mathrm{(a)}$ & $4.6$ $^\mathrm{(a)}$ & $4.9$ $^\mathrm{(a)}$ \\
$kT_1$ ($\mathrm{keV}$) & $0.674^{+0.006}_{-0.005}$ & $0.653^{+0.009}_{-0.009}$ & $0.638^{+0.011}_{-0.018}$ \\
$R_{\mathrm{BB}_1}^\mathrm{(b)}$ ($\mathrm{km}$) & $1.20^{+0.02}_{-0.02}$ & $1.27^{+0.04}_{-0.03}$ & $1.36^{+0.06}_{-0.05}$ \\
$kT_2$ ($\mathrm{keV}$) & --- & $<200$ & --- \\
$R_{\mathrm{BB}_2}^\mathrm{(b)}$ ($\mathrm{km}$) & --- & $1.91^{+182.86}_{-0.20}\times10^{-3}$ & --- \\
$\Gamma_\mathrm{PL}$ & --- & --- & $-1.14^{+2.81}_{-1.10}$ \\
$\mathrm{norm}_\mathrm{PL}^\mathrm{(c)}$ & --- & --- & $3.46^{+571.08}_{-3.21}\times10^{-6}$ \\
$F_\mathrm{unabs}^{2-6}$$^\mathrm{(d)}$ & $9.29^{+0.07}_{-0.07}$ & $9.37^{+0.07}_{-0.07}$ & $9.69^{+0.07}_{-0.07}$ \\
$F_\mathrm{obs}^{2-6}$ $^\mathrm{(e)}$ & $5.81^{+0.04}_{-0.04}$ & $5.87^{+0.04}_{-0.04}$ & $5.87^{+0.04}_{-0.04}$ \\
$\mathrm{const}_{\mathrm{DU}_1}$ & $1.00$ $^\mathrm{(a)}$ & $1.00$ $^\mathrm{(a)}$ & $1.00$ $^\mathrm{(a)}$ \\
$\mathrm{const}_{\mathrm{DU}_2}$ & $1.02^{+0.02}_{-0.02}$ & $1.03^{+0.02}_{-0.02}$ & $1.02^{+0.02}_{-0.02}$ \\
$\mathrm{const}_{\mathrm{DU}_3}$ & $0.95^{+0.02}_{-0.02}$ & $0.95^{+0.02}_{-0.02}$ & $0.95^{+0.02}_{-0.02}$ \\
$\chi^2/\mathrm{dof}$ & $263.95/244$ & $255.50/242$ & $253.94/242$ \\
\hline\hline
\end{tabular}
    \endgroup
    \end{center}
    \tablecomments{ \\
    $\mathrm{(a)}$ -- Frozen parameters.\\
    $\mathrm{(b)}$ -- Emission radius extracted from the \texttt{bbodyrad} parameter $\mathrm{norm}$ assuming a distance of $4.5\,\mathrm{kpc}$ \cite[][]{2010ApJ...710..227T}. \\
    $\mathrm{(c)}$  -- Normalization of the \texttt{powerlaw} component in $\mathrm{photons\,keV^{-1}\,cm^{-2}\,s^{-1}}$ at $1\,\mathrm{keV}$. \\
    $\mathrm{(d)}$ -- Total unabsorbed $2$--$6\,\mathrm{keV}$ flux in units of $10^{-12}\,\mathrm{erg\,cm^{-2}\,s^{-1}}$. \\
    $\mathrm{(e)}$ -- Total observed $2$--$6\,\mathrm{keV}$ flux in units of $10^{-12}\,\mathrm{erg\,cm^{-2}\,s^{-1}}$.}
\end{table*}

\subsection{Phase-resolved spectral analysis} \label{sec:spectral_pr}
\begin{figure}[]
\includegraphics[width=0.47\textwidth]{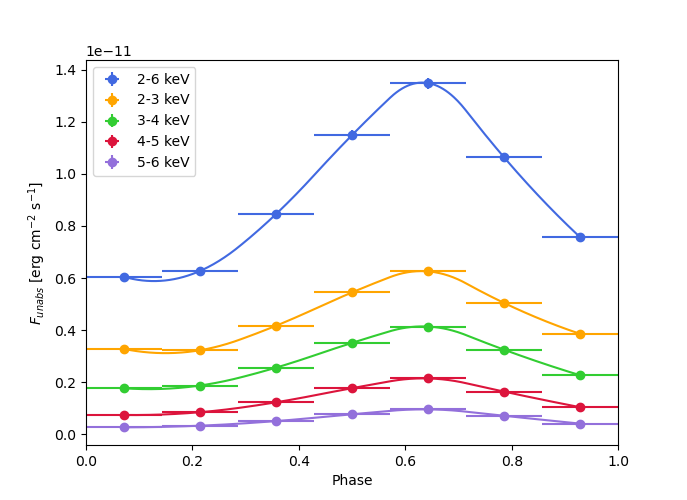}
\caption{\ixpe\ pulse-profile of \source\ in the  $2$--$6$, $2$--$3$, $3$--$4$, $4$--$5$ and $5$--$6\,\mathrm{keV}$ energy bands; the solid lines are splines connecting the data. 
\label{fig:pp_pcube_26}}
\end{figure}

Using the \textsc{ixpeobssim} \texttt{xpphase} and \texttt{xpselect} tools, the data were folded into seven equally-spaced phase bins, based on the timing solution presented in \S\ref{sec:timing}. We then extracted the source and background spectra in each phase bin, 
and, according to the results of the phase-averaged analysis discussed in \S\ref{sec:spectral-pa}, we fit the data using a single-component model (\texttt{tbabs}$\times$\texttt{bbodyrad}), with the column density $N_\mathrm{H}$ frozen to the value reported in Table \ref{tab:spec_xspec}\footnote{We verified that, similarly to what was found in the phase-averaged spectral analysis, a second spectral component is not statistically required.}. 

\begin{table*}[ht!]
\tabletypesize{\scriptsize}
\begin{center}
\caption{Phase-resolved spectral fit of the \ixpe\ data in the $2$--$6\,\mathrm{keV}$ band. \label{tab:spec_xspec_pr}}
\begingroup
\renewcommand{\arraystretch}{1.2}
\begin{tabular}{c}
Spectral model: \texttt{tbabs}$\times$\texttt{bbodyrad}$^\mathrm{(a)}$ \\
\end{tabular}
\setlength{\tabcolsep}{22pt}
\begin{tabular}{c c c c c c}
\hline\hline
phase bin & $kT$ ($\mathrm{keV}$) & $R_\mathrm{BB}$ ($\mathrm{km}$)$^\mathrm{(b)}$ & $F_\mathrm{unabs}^\mathrm{(c)}$ & $F_\mathrm{obs}^\mathrm{(d)}$ & $\chi^2/\mathrm{dof}$ \\
\hline
$0.00$--$0.14$ & $0.610^{+0.018}_{-0.017}$ & $1.24^{+0.24}_{-0.23}$ & $6.05^{+0.15}_{-0.15}$ & $3.63^{+0.09}_{-0.09}$ & $127.00/131$ \\
$0.14$--$0.29$ & $0.641^{+0.018}_{-0.017}$ & $1.11^{+0.21}_{-0.20}$ & $6.27^{+0.15}_{-0.15}$ & $3.84^{+0.09}_{-0.09}$ & $103.07/128$ \\
$0.29$--$0.43$ & $0.667^{+0.017}_{-0.017}$ & $1.17^{+0.21}_{-0.19}$ & $8.46^{+0.17}_{-0.17}$ & $5.26^{+0.11}_{-0.11}$ & $128.68/149$ \\
$0.43$--$0.57$ & $0.691^{+0.014}_{-0.014}$ & $1.25^{+0.20}_{-0.19}$ & $11.49^{+0.20}_{-0.20}$ & $7.25^{+0.12}_{-0.12}$ & $159.02/164$ \\
$0.57$--$0.71$ & $0.704^{+0.014}_{-0.013}$ & $1.29^{+0.19}_{-0.18}$ & $13.49^{+0.22}_{-0.22}$ & $8.57^{+0.14}_{-0.14}$ & $164.24/178$ \\
$0.71$--$0.86$ & $0.690^{+0.015}_{-0.015}$ & $1.21^{+0.19}_{-0.18}$ & $10.63^{+0.19}_{-0.19}$ & $6.71^{+0.12}_{-0.12}$ & $168.32/161$ \\
$0.86$--$1.00$ & $0.646^{+0.017}_{-0.017}$ & $1.20^{+0.22}_{-0.20}$ & $7.56^{+0.16}_{-0.16}$ & $4.65^{+0.10}_{-0.10}$ & $148.12/137$ \\
\hline\hline
\end{tabular}
    \endgroup
    \end{center}
    \tablecomments{\\
    (a) -- $N_\mathrm{H}$ is fixed at $4.6\times10^{22}\,\mathrm{cm}^{-2}$ \cite[see][]{2020A&A...633A..31C} and the cross-calibration constants of the three \ixpe\ DUs are frozen to the values reported in Table \ref{tab:spec_xspec}. \\
    (b) -- Emission radius extracted from the \texttt{bbodyrad} parameter $\mathrm{norm}$ assuming a distance of $4.5\,\mathrm{kpc}$ \cite[][]{2010ApJ...710..227T}.\\
    (c) -- Total unabsorbed flux in units of $10^{-12}\,\mathrm{erg\,cm^{-2}\,s^{-1}}$. \\
    (d) -- Total observed flux in units of $10^{-12}\,\mathrm{erg\,cm^{-2}\,s^{-1}}$.}
\end{table*}

Table \ref{tab:spec_xspec_pr} reports the best-fitting parameters for each phase bin, together with the corresponding  unabsorbed and observed fluxes. The $\chi^2/\mathrm{dof}$ values indicate that the model provides an acceptable fit at a confidence level greater than $99\%$ in all phase bins. The pulse-profile in the $2$--$6\,\mathrm{keV}$ energy range is shown in Figure \ref{fig:pp_pcube_26}, along with those in the $2$--$3$, $3$--$4$, $4$--$5$ and $5$--$6\,\mathrm{keV}$ energy bins. The average values of the unabsorbed and observed fluxes, $(9.14\pm0.07)\times10^{-12}\,\mathrm{erg\,cm^{-2}\,s^{-1}}$ and $(5.70\pm0.04)\times10^{-12}\,\mathrm{erg\,cm^{-2}\,s^{-1}}$, respectively, are consistent with the corresponding phase-averaged values (see Table \ref{tab:spec_xspec}) at $2\sigma$ confidence level. 
The measured pulsed fraction (PF, see Table \ref{tab:pdpa_xspec}) increases monotonically with energy from $31.9\pm1.3\%$ at $2$--$3\,\mathrm{keV}$ to $56.5\pm1.0\%$ at $5$--$6\,\mathrm{keV}$ (the constant-PF hypothesis is rejected at more than $10\sigma$). The averaged pulsed fraction in the $2$--$6\,\mathrm{keV}$ is $38.1\pm1.3\,\%$ .

\begin{figure}[]
\includegraphics[width=0.47\textwidth]{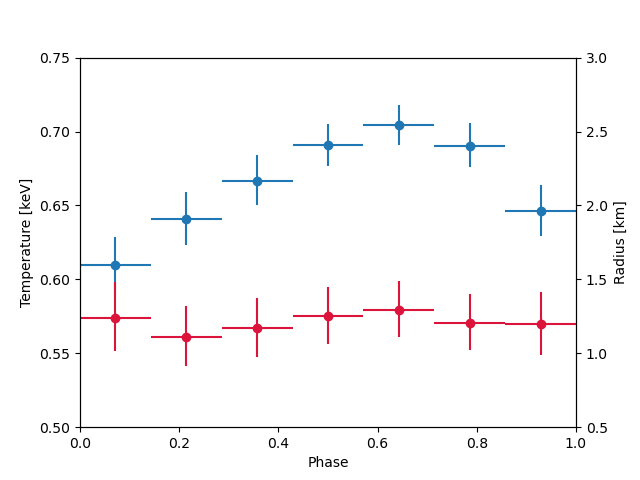}
\caption{Phase variation of the blackbody temperature (cyan) and radius (red) for the best-fitting model reported in Table \ref{tab:spec_xspec_pr}.
\label{fig:param_spec_pr}}
\end{figure}
Figure \ref{fig:param_spec_pr} shows the phase dependence of the temperature and emission radius for the best-fitting spectral model. While $R_\mathrm{BB}$ is consistent with a constant value ($1.22\pm0.07\,\mathrm{km}$) at the $1\sigma$ confidence level, in agreement with the emission radius inferred from the phase-averaged spectrum (see Table \ref{tab:spec_xspec}), the blackbody temperature shows a clear oscillation in phase with the pulse profile (see Figure \ref{fig:pp_pcube_26}). 

\begin{figure}[h]
\includegraphics[width=0.47\textwidth]{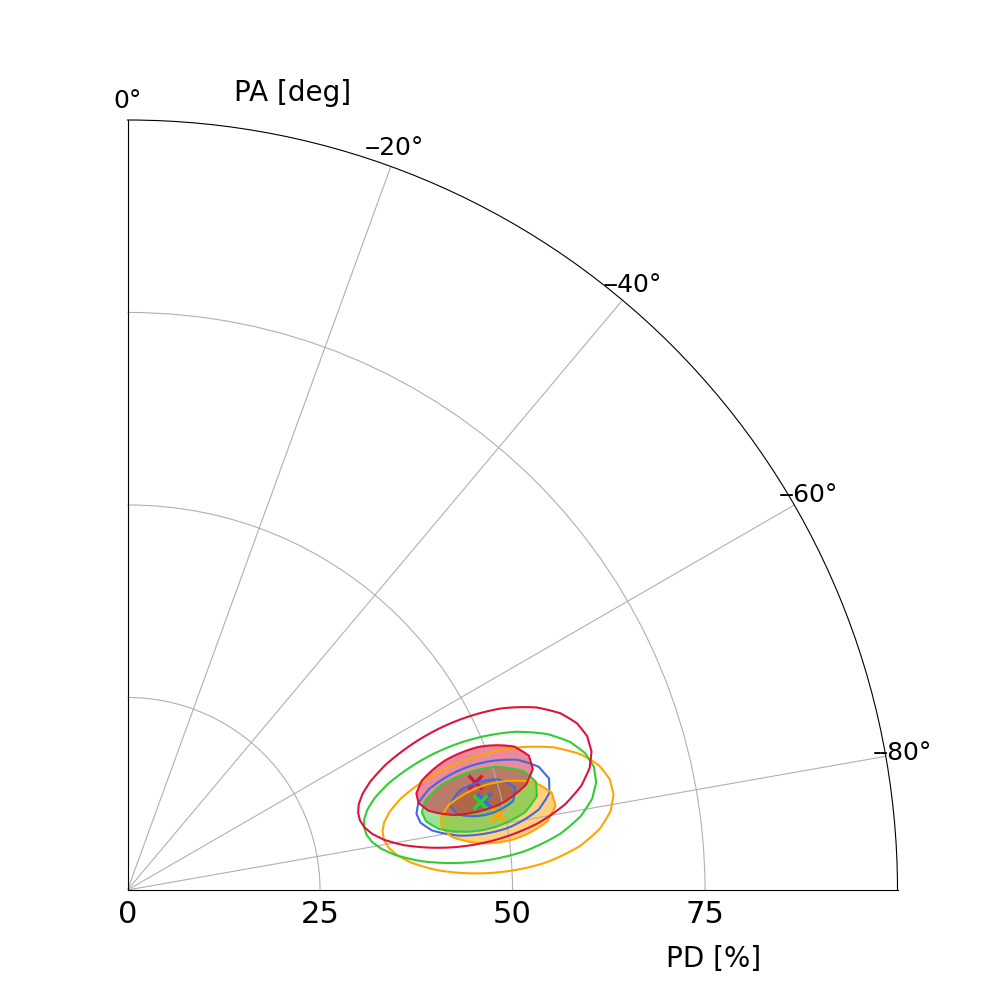}
\caption{\source\ phase- and energy-integrated ($2$--$6\,\mathrm{keV}$) most-probable values for $\mathrm{PD}$ and $\mathrm{PA}$ (crosses) measured by the three \ixpe\ DUs (DU1 yellow, DU2 green and DU3 red); the combined result is also shown in blue. Constant PD and constant PA loci are marked by concentric circles and radial lines, respectively, with $\mathrm{PA}=0^\circ$ corresponding to the celestial North, and PA decreasing Westward. The shaded and empty contours represents the $68\%$ and $99\%$ confidence regions, respectively, obtained using the \texttt{steppar} command in \textsc{xspec}.
\label{fig:mirino26}}
\end{figure}
\subsection{Phase-averaged polarimetric analysis} \label{sec:polarimetric_pa}
We processed the source photon lists using the same procedure described in \S\ref{sec:spectral-pa} and \S\ref{sec:spectral_pr} to extract the $Q$ and $U$ Stokes parameters as functions of photon energy and rotational phase. 
We followed the method discussed in \citet{2017ApJ...838...72S}, simultaneously fitting the $I$, $Q$, and $U$ spectra in different energy intervals within \textsc{xspec}, with the appropriate response functions; the $Q$ and $U$ data were grouped in such a way to ensure at least $5$ counts per energy bin. We convolved a constant polarization model (\texttt{polconst}) to the \texttt{tbabs}$\times$\texttt{bbodyrad}  model (see \S\ref{sec:spectral-pa}) and froze the spectral parameters to the values obtained in the phase-averaged spectral fit (see Table \ref{tab:spec_xspec}). We note that this method is essentially independent of the specific spectral model adopted, provided that the energy intervals in which the polarization properties are extracted are sufficiently narrow. We checked a posteriori that the PD and PA values derived using the alternative spectral models reported in Table \ref{tab:spec_xspec} differ only marginally from those obtained using the single-BB model, with deviations well within the $1\sigma$ confidence intervals. 

The results for the phase-averaged degree and angle of polarization, integrated in the entire range $2$--$6\,\mathrm{keV}$ are shown in Figure \ref{fig:mirino26} for each of the three DUs, together with the value obtained by combining the data from all DUs. The polarization degree for the combined DUs, $\mathrm{PD}=\sqrt{(Q/I)^2+(U/I)^2}$, is $47.7\pm2.9\%$ and is significant at the level $\approx16\sigma$. 
The associated polarization angle, $\mathrm{PA}=\arctan(U/Q)/2$ is $75^\circ.8\pm1^\circ.8$, measured West of North. The values of $\mathrm{PD}$ and $\mathrm{PA}$ measured by the individual DUs 
differ by less than $1\sigma$, and are entirely consistent with each other. All measurements are above $\mathrm{MDP}_{99}$ \footnote{The minimum detectable polarization at the $99\%$ confidence level as defined in \citet{2010ApJ...713..912W}.}, as certified by the fact that the $99\%$ confidence contours in the $\mathrm{PD}$--$\mathrm{PA}$ polar plane are closed.

\begin{figure}[t]
\includegraphics[width=0.47\textwidth]{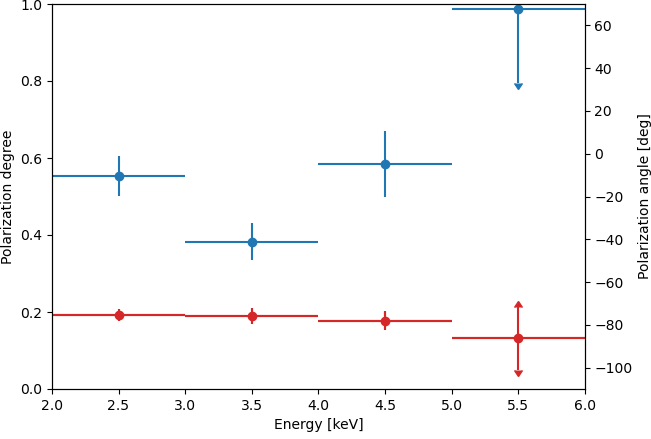}
\includegraphics[width=0.47\textwidth]{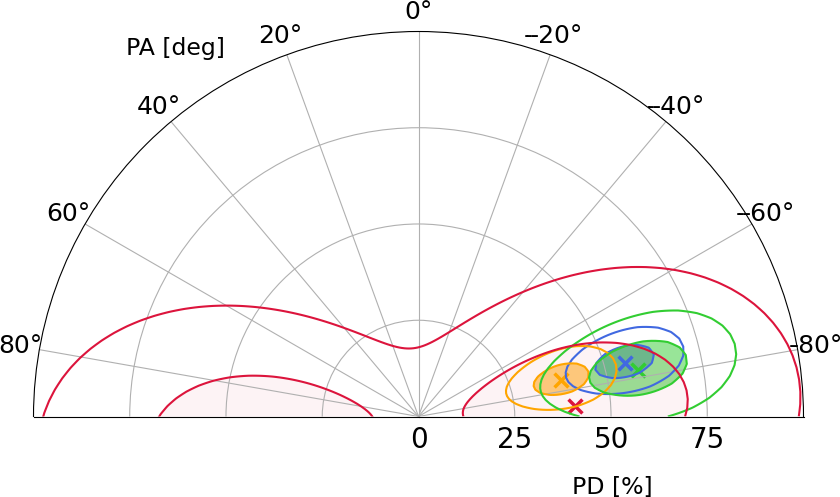}
\caption{Top: energy-dependent polarization degree (cyan) and polarization angle (red) obtained with \textsc{xspec} (see text for details). Error bars show the $1\sigma$ confidence intervals derived with the \texttt{err} procedure. When the PD measurement falls below $\mathrm{MDP}_{99}$, a downwards arrow  marks the $3\sigma$ upper limit; the associated value of PA is unconstrained  and is shown by a double-headed arrow. Bottom: same in the PD-PA plane for the $2$--$3$ (blue), $3$--$4$ (orange), $4$--$5$ (green) and $5$--$6\,\mathrm{keV}$ (red) energy bins; symbols and contour codes are as in  Figure \ref{fig:mirino26}.
All values are reported in Table \ref{tab:pdpa_xspec}.\label{fig:pdpa_xspec}}
\end{figure}

The energy-dependent degree and angle of polarization in the range $2$--$6\,\mathrm{keV}$ are shown in Figure \ref{fig:pdpa_xspec}, while the best-fit values obtained in the different energy intervals are reported in Table \ref{tab:pdpa_xspec} (along with the corresponding pulsed fraction). When the most probable value of PD in a given energy interval falls below the corresponding $\mathrm{MDP}_{99}$ (i.e., when the corresponding $99\%$ confidence contour remains open in the bottom panel of Figure \ref{fig:pdpa_xspec}), the $3\sigma$ upper limit is reported in the top panel of Figure \ref{fig:pdpa_xspec}, while PA is unconstrained and marked by a double arrow. Significant polarization is detected only at energies below $5\,\mathrm{keV}$, which is in agreement with the discussion in section \ref{sec:obs-ixpe}. In bins where the polarization measure is significant, the PD appears to vary with energy, showing a minimum between $3$ and $4\,\mathrm{keV}$. Taking into account also the most probable value of $\mathrm{PD}$ in the bin $5$--$6\,\mathrm{keV}$, the hypothesis of a constant $\mathrm{PD}$ is marginally rejected at the confidence level of $94.9\%$ ($\chi^2/\mathrm{dof}=7.75/3$, for a fit to a constant $\mathrm{PD}$). Considering, instead, only the 3 energy bins where the polarization degree is significant, the constant fit is ruled out at $97.6\%$ confidence level ($\chi^2/\mathrm{dof}=7.48/2$). This is supported by the confidence contours shown in Figure \ref{fig:pdpa_xspec}: the $68\%$ contours of the $3$--$4\,\mathrm{keV}$ and the $2$--$3\,\mathrm{keV}$ energy bins do not intersect and the former is only tangent to the $4$--$5\,\mathrm{keV}$ one. The polarization angle, instead, is fully consistent with a constant within $1\sigma$ confidence level, with an average value of $74^\circ.4\pm2^\circ.1$ measured West of North.

\begin{table*}[ht!]
\tabletypesize{\scriptsize}
\begin{center}
\caption{Energy-dependent pulsed fraction, polarization degree and angle. \label{tab:pdpa_xspec}}
\begingroup
\setlength{\tabcolsep}{32.pt}
\renewcommand{\arraystretch}{1.5}
\begin{tabular}{c c c c c }
\hline\hline
Energy bin & PF [$\%$] & PD [$\%$] & PA [deg] & $\chi^2/\mathrm{dof}$ \\
\hline
$2$--$3\,\mathrm{keV}$ & $31.9^{+1.3}_{-1.3}$ & $55.4^{+5.2}_{-5.2}$ & $-75.4^{+2.7}_{-2.7}$ & $97.95/94$ \\
$3$--$4\,\mathrm{keV}$ & $40.1^{+1.2}_{-1.2}$ & $38.3^{+4.9}_{-4.9}$ & $-75.7^{+3.7}_{-3.7}$ & $82.09/94$ \\
$4$--$5\,\mathrm{keV}$ & $48.7^{+1.1}_{-1.1}$ & $58.3^{+8.6}_{-8.6}$ & $-78.0^{+4.3}_{-4.3}$ & $114.52/93$ \\
$5$--$6\,\mathrm{keV}$ & $56.5^{+1.0}_{-1.0}$ & ($40.8^{+19.3}_{-19.3}$) & ($-86.2^{+176.2}_{-3.8}$) & $50.12/46$ \\
\hline\hline
\end{tabular}
    \endgroup
    \end{center}
    \tablecomments{PF, PD and PA values are obtained using the \textsc{xspec} procedure described in the text, with errors (at $1\sigma$ confidence level) derived from the \texttt{err} procedure in \textsc{xspec}. Negative $\mathrm{PA}$ values are given West of the celestial North. The most probable values of $\mathrm{PD}$ and $\mathrm{PA}$ are reported in parenthesis when PD falls below the corresponding $\mathrm{MDP}_{99}$.}
\end{table*}

\subsection{Phase-resolved polarimetric analysis}\label{sec:phaseres}

In order to investigate how the polarization properties depend on the rotational phase, we folded the source photon lists into seven equally-spaced phase bins as described in \S\ref{sec:spectral_pr}, i.e. through the \textsc{ixpeobssim} tools \texttt{xpphase} and \texttt{xpselect} and using the timing solution presented in \S\ref{sec:timing}. We then performed a simultaneous fit of the Stokes spectra within \textsc{xspec}, using a \texttt{tbabs}$\times$\texttt{(bbodyrad}$\times$\texttt{polconst}) model, with all spectral parameters frozen at those reported in Table \ref{tab:spec_xspec_pr} in the different phase bins and leaving only the \texttt{polconst} $\mathrm{PD}$ and $\mathrm{PA}$ as free parameters. The values of the best fitting parameters are reported in Table \ref{tab:pdpa_xspec_pr}.

\begin{figure}[h]
\includegraphics[width=0.47\textwidth]{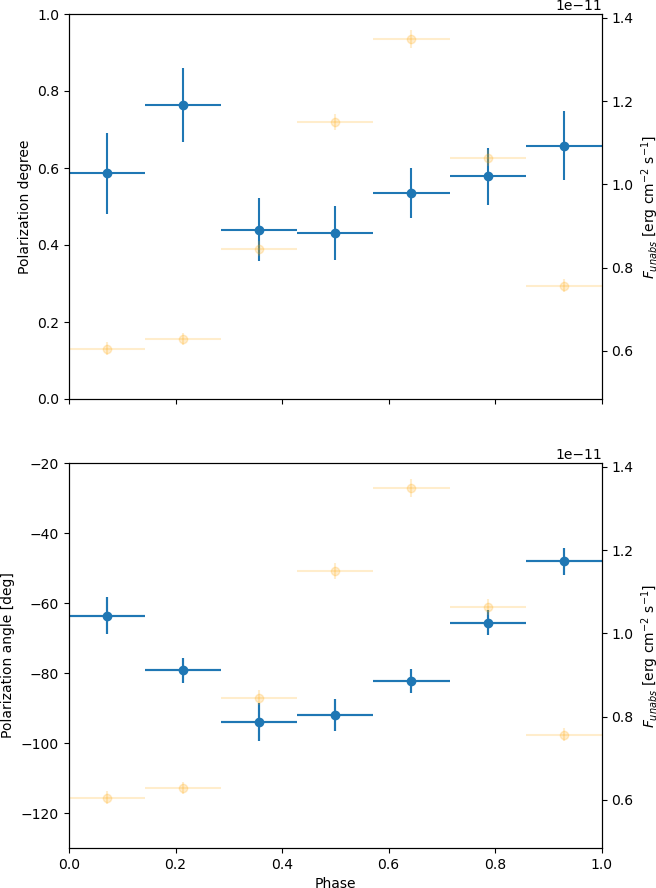}
\caption{Phase-dependent polarization degree (top) and polarization angle (bottom), integrated over the $2$--$6\,\mathrm{keV}$ range, obtained using  \textsc{xspec} (cyan, see text for details). Errors  are at $1\sigma$, as  derived with the \textsc{xspec} \texttt{err} procedure. Values are reported in Table \ref{tab:pdpa_xspec_pr}.  
In both plots, the $2$--$6\,\mathrm{keV}$ pulse profile has been superimposed for ease of comparison (light orange). A shift by $-180^\circ$ was applied to all positive values of $\mathrm{PA}$ (third and fourth phase bins, see Table \ref{tab:pdpa_xspec_pr}) to reproduce a smooth oscillation.
\label{fig:pdpa_xspec_pr}}
\end{figure}

Figure \ref{fig:pdpa_xspec_pr} illustrates the phase-dependent behavior of $\mathrm{PD}$ and $\mathrm{PA}$ in the range $2$--$6\,\mathrm{keV}$, with the corresponding pulse profile (see Figure \ref{fig:pp_pcube_26}) overlayed for comparison.  $\mathrm{PD}$ is only marginally consistent with a constant ($\chi^2/\mathrm{dof}=9.38/6$, $p$-value of $0.153$), and  the most probable values hint at an anti-correlation with the pulse profile, with the minimum (maximum) of PD occurring in the vicinity of the maximum (minimum) of the pulse profile. In contrast, the polarization angle exhibits a clear oscillation with the rotational phase, around an average value $-74^\circ.3\pm1^\circ.5$, consistent (within $1\sigma$) with that derived in \S\ref{sec:polarimetric_pa} (see Figure \ref{fig:mirino26}).

\begin{figure}[h]
\includegraphics[width=0.47\textwidth]{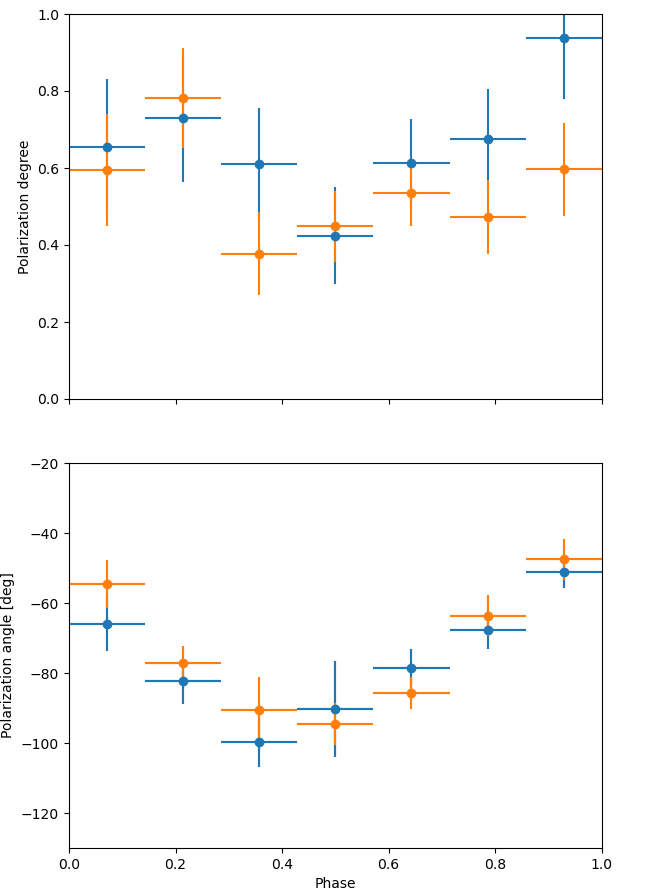}
\caption{Phase-dependent polarization degree (top) and polarization angle (bottom), integrated over energy in the $2$--$3$ (blue) and $3$--$6\,\mathrm{keV}$ (orange) ranges (errors are at $1\sigma$ confidence level; the error bar of $\mathrm{PD}$ in the last bin of the top panel is truncated at the maximum attainable value, $1$). Details as in Figure \ref{fig:pdpa_xspec_pr}. Values are reported in Table \ref{tab:pdpa_xspec_pr}. 
\label{fig:pdpa_xspec_pr_2336}}
\end{figure}

\begin{table*}[ht!]
\tabletypesize{\scriptsize}
\begin{center}
\caption{Phase-resolved polarization analysis. \label{tab:pdpa_xspec_pr}}
\begingroup
\renewcommand{\arraystretch}{1.2}
\begin{tabular}{c}
$2$--$6\,\mathrm{keV}$ \\
\end{tabular}
\setlength{\tabcolsep}{37.pt}
\begin{tabular}{c c c c}
\hline\hline
phase bin & $\mathrm{PD}\,[\%]$ & $\mathrm{PA}\,[\mathrm{deg}]$ & $\chi^2/\mathrm{dof}$ \\
\hline
$0.00$--$0.14$ & $58.6^{+10.5}_{-10.5}$ & $-63.6^{+5.2}_{-5.1}$ & $228.53/247$ \\
$0.14$--$0.29$ & $76.4^{+9.7}_{-9.7}$ & $-79.2^{+3.6}_{-3.6}$ & $225.15/244$ \\
$0.29$--$0.43$ & $44.0^{+8.2}_{-8.2}$ & $86.0^{+5.4}_{-5.4}$ & $240.34/265$ \\
$0.43$--$0.57$ & $43.1^{+7.0}_{-7.0}$ & $88.0^{+4.7}_{-4.7}$ & $273.30/280$ \\
$0.57$--$0.71$ & $53.5^{+6.4}_{-6.4}$ & $-82.3^{+3.4}_{-3.4}$ & $258.64/294$ \\
$0.71$--$0.86$ & $57.8^{+7.3}_{-7.3}$ & $-65.6^{+3.6}_{-3.6}$ & $276.18/277$ \\
$0.86$--$1.00$ & $65.8^{+9.0}_{-9.0}$ & $-48.1^{+3.9}_{-3.9}$ & $272.99/253$ \\
\hline\hline
\end{tabular}
\begin{tabular}{c}
$2$--$3\,\mathrm{keV}$ \\
\end{tabular}
\setlength{\tabcolsep}{37.5pt}
\begin{tabular}{c c c c}
\hline\hline
phase bin & $\mathrm{PD}\,[\%]$ & $\mathrm{PA}\,[\mathrm{deg}]$ & $\chi^2/\mathrm{dof}$ \\
\hline
$0.00$--$0.14$ & $65.5^{+17.6}_{-17.6}$ & $-66.0^{+7.7}_{-7.9}$ & $87.99/94$ \\
$0.14$--$0.29$ & $73.1^{+16.7}_{-16.7}$ & $-82.4^{+6.4}_{-6.4}$ & $80.74/91$ \\
$0.29$--$0.43$ & $61.0^{+14.6}_{-14.6}$ & $-80.2^{+6.9}_{-6.9}$ & $83.12/94$ \\
$0.43$--$0.57$ & $42.4^{+12.6}_{-12.6}$ & $89.7^{+8.6}_{-8.6}$ & $85.05/94$ \\
$0.57$--$0.71$ & $61.2^{+11.6}_{-11.6}$ & $-78.5^{+5.5}_{-5.5}$ & $87.49/94$ \\
$0.71$--$0.86$ & $67.5^{+12.9}_{-12.9}$ & $-67.7^{+5.5}_{-5.5}$ & $103.25/94$ \\
$0.86$--$1.00$ & $93.9^{+6.1}_{-15.9}$ & $-51.1^{+4.7}_{-4.7}$ & $88.07/94$ \\
\hline\hline
\end{tabular}
\begin{tabular}{c}
$3$--$6\,\mathrm{keV}$ \\
\end{tabular}
\setlength{\tabcolsep}{37.pt}
\begin{tabular}{c c c c}
\hline\hline
phase bin & $\mathrm{PD}\,[\%]$ & $\mathrm{PA}\,[\mathrm{deg}]$ & $\chi^2/\mathrm{dof}$ \\
\hline
$0.00$--$0.14$ & $59.5^{+14.5}_{-14.5}$ & $-54.7^{+7.0}_{-7.0}$ & $129.14/142$ \\
$0.14$--$0.29$ & $78.1^{+13.1}_{-13.1}$ & $-77.2^{+4.8}_{-4.8}$ & $139.02/142$ \\
$0.29$--$0.43$ & $37.7^{+10.8}_{-10.8}$ & $89.3^{+8.5}_{-8.5}$ & $150.40/160$ \\
$0.43$--$0.57$ & $44.9^{+9.3}_{-9.3}$ & $85.4^{+6.0}_{-6.0}$ & $181.47/175$ \\
$0.57$--$0.71$ & $53.4^{+8.4}_{-8.4}$ & $-85.7^{+4.5}_{-4.5}$ & $164.68/189$ \\
$0.71$--$0.86$ & $47.3^{+9.6}_{-9.6}$ & $-63.6^{+5.8}_{-5.9}$ & $157.96/172$ \\
$0.86$--$1.00$ & $59.6^{+12.1}_{-12.1}$ & $-47.5^{+5.9}_{-5.9}$ & $172.56/148$ \\
\hline\hline
\end{tabular}
\endgroup
\end{center}
    \tablecomments{Results are obtained using the \textsc{xspec} procedure described in the text. Errors are at $1\sigma$ confidence level, derived from the \texttt{err} procedure in \textsc{xspec}. $\mathrm{PA}$ values are given East of the celestial North. 
    }
\end{table*}

We further analyzed the data by splitting the $2$--$6\,\mathrm{keV}$ range into narrower energy intervals. After exploring different binning schemes, we found that a significant detection ($\mathrm{PD}>\mathrm{MDP}_{99}$) is obtained in all phase intervals only in the $2$--$3\,\mathrm{keV}$ and $3$--$6\,\mathrm{keV}$ bands. 
As shown in Figure \ref{fig:pdpa_xspec_pr_2336}, the phase-dependent polarization degree changes little in the two bands (a difference of more  than $1\sigma$ appears only in the last phase bin). On the other hand, the polarization angle displays a clear oscillation in both the $2$--$3$ and the $3$--$6\,\mathrm{keV}$ bands, consistent with the behavior in the total energy range.

\begin{figure}[t]
\includegraphics[width=0.47\textwidth]{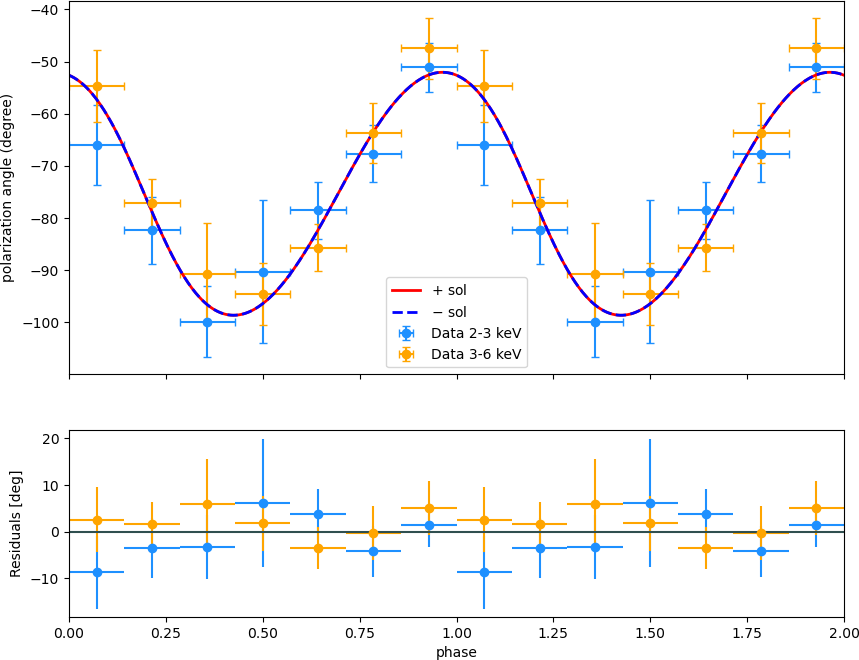}
\caption{\ixpe\ $2$--$3\,\mathrm{keV}$ (cyan) and $3$--$6\,\mathrm{keV}$  (orange) polarization angle simultaneously fitted by a rotating vector model (see Table \ref{tab:rvm_fit}); two rotational cycles are shown for ease of visualization. The red solid (blue dotted)  line shows the best fitting plus (minus) solution; residuals are shown in the bottom panel.
\label{fig:rvm_sim_fit}}
\end{figure}

\begin{figure}[t]
\includegraphics[width=0.47\textwidth]{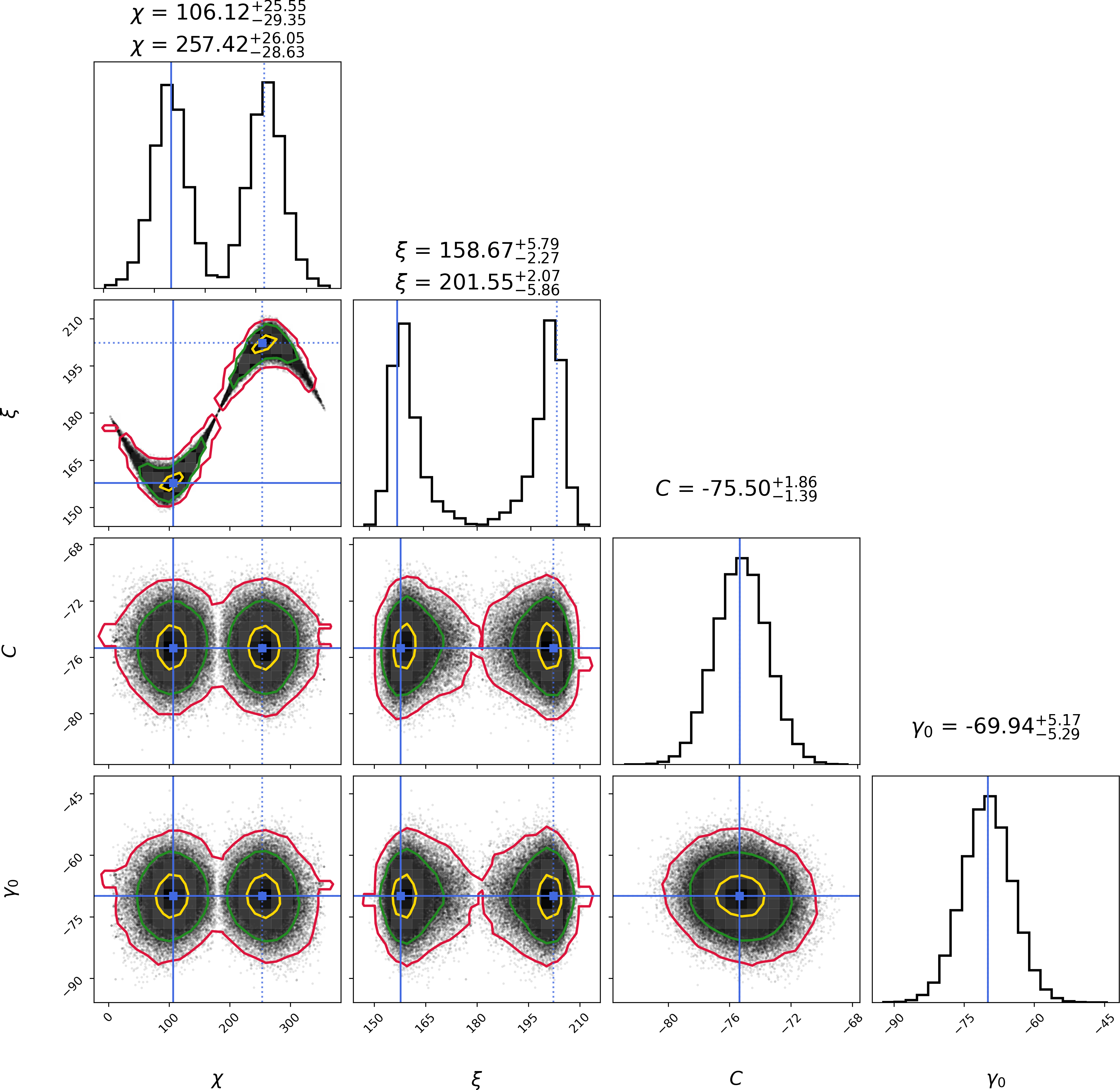}
\\
\\
\includegraphics[width=0.47\textwidth]{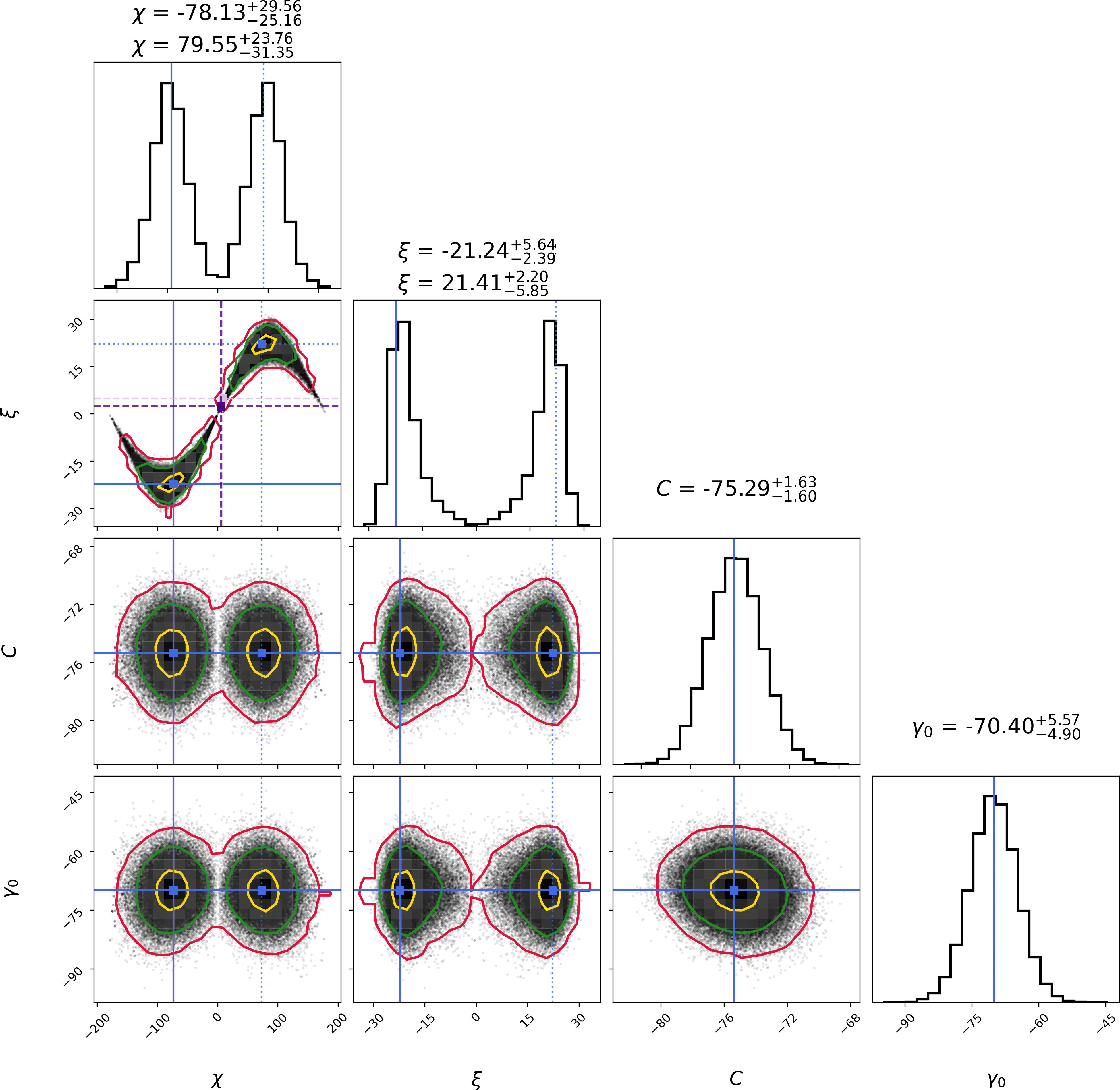}
\caption{Marginalized posterior distributions for the RVM plus (top) and minus (bottom) solutions. The best-fit values ($\chi_\pm,\xi_\pm$), obtained through $\chi^2$ minimization, as well as the mirrored solutions ($2\pi-\chi_\pm,2\pi-\xi_\pm$) are marked by solid and dotted blue lines, respectively. The 1, 2, and 3$\sigma$ confidence contours are shown in yellow, green, and red, respectively. The maximum a posteriori estimates, along with their associated $1\sigma$ uncertainties, are reported above each 1D histogram. The best-fit values from the radio analysis by \citet{2025arXiv250919446S}, $\chi=5^\circ.2,\xi=2^\circ.4$ (dashed violet lines) and $\chi=8^\circ.4,\xi=4^\circ.9$ (dashed pink lines), are also shown for comparison.
\label{fig:post}}
\end{figure}

We fitted the phase-dependent polarization angle with the rotating vector model \cite[RVM,][see \citealt{2021MNRAS.502.1549T} and \citealt{taverna+22} for a discussion on the RVM model applied to magnetars]{1969ApL.....3..225R,1970Natur.225..612K},
\begin{equation}
    \tan(\mathrm{PA}-C) = \frac{\sin\xi\sin[\pm(\gamma+\gamma_0)]}{\sin\chi\cos\xi-\cos\chi\sin\xi\cos(\gamma+\gamma_0)},
    \label{eqn:rvm}
\end{equation}
where $\chi$ and $\xi$ are the angles of the line-of-sight (LOS) and of the magnetic axis with the spin axis, respectively, $\gamma$ is the rotational phase, $\gamma_0$ is an arbitrary initial phase, and $C$ is an angular offset. The plus/minus sign refers to the two possible directions of the star rotation, counter-clockwise or clockwise, and reflects the two conventions commonly adopted in the literature about the direction in which the position angle increases in the plane of the sky, clockwise/counter-clockwise if a plus/minus sign is placed in front of the fraction in equation (\ref{eqn:rvm}), in lieu of the plus/minus sign inside the sine function \citep[][]{2001ApJ...553..341E}. For the sake of conciseness, we will refer to the two possible choices for the RVM models as the ``plus'' ($\chi_+,\xi_+$) and ``minus'' ($\chi_-,\xi_-$) solutions. It should be noted that, due to the $\pi$-periodicity of the $\arctan$ function, the configurations $(2\pi-\chi_\pm,2\pi-\xi_\pm)$, i.e. $(-\chi_\pm,-\xi_\pm)$, describe equivalent geometries.
In order to improve the statistics, we fitted simultaneously the data integrated in the $2$--$3$ and in the $3$--$6\,\mathrm{keV}$ band, treating the two datasets as mutually independent. This is justified on the assumption that the recorded events are Poissonian distributed and the events collected in separate energy channels are physically distinct. 
Leaving all four parameters in equation (\ref{eqn:rvm}) free to vary, the best simultaneous fits yield $\chi_+=106^\circ.6\pm30^\circ.9$ and $\xi_+=157^\circ.7\pm4^\circ.0$ for the plus solution and $\chi_-=-73^\circ.4,\,\xi_-=-22^\circ.3$ for the minus one (see Table \ref{tab:rvm_fit} and Figure \ref{fig:rvm_sim_fit} for more details). Both pairs of angles provide the same, good fit to the observed $\mathrm{PA}$ and, as expected, it is $\xi_-=\xi_+-\pi,\, \chi_-=\chi_+-\pi$, which exactly mirrors the two spin orientations. We also verified that the fits of $\mathrm{PA}$ integrated throughout the $2$--$6\,\mathrm{keV}$ band yield consistent results within the errors.
We corroborated our RVM results with a Bayesian analysis, starting from flat priors in the range $[0,\,2\pi]$ for the plus solution, 
$[-\pi,\,\pi]$ for the minus solution, 
$[-2\pi,\,2\pi]$ for $C$ and $\gamma_0$. The posteriors for the two models are shown in Figure \ref{fig:post}; the probability density distributions are clearly the same, much as the best fitting models in Figure \ref{fig:rvm_sim_fit}, with the only difference of a shift by $-\pi$ in the $\chi$, $\xi$ angles. The best estimates for the parameters derived with both techniques are in close agreement, well within $1\sigma$ uncertainties.

Although the RVM fitting provides a well constrained value of the angle $\xi$ between the rotation and the dipole axis, the angle $\chi$ between the rotation axis and the LOS is  more loosely determined, with an uncertainty of $\sim 30^\circ$. Despite these limitations, both solutions point to an inclined rotator seen away from the spin axis. This is at variance with the  geometry reported by \citet{2025arXiv250919446S}, $\chi=5^\circ.2^{+3.2}_{-2.6}$ and $\xi=2^\circ.4^{+1.4}_{-1.2}$, based on simultaneous Parkes/Murriyang radio data. 
Our solution for the inclination of the dipole axis, $\xi_+\sim 158^\circ$, is close to the best estimate, $\xi=160^\circ$, reported by \citet{2008ApJ...679..681C}, using two distinct 2007 Parkes radio observations covering roughly half of the rotational period. However, the value they derived for the inclination of the spin axis, $\chi=174^\circ$, is different. We also note that a geometry with ($\chi=174^\circ,\, \xi=160^\circ$) is inconsistent with the oscillating behavior of $\mathrm{PA}$ observed in the radio, since the RVM model with $\chi>\xi$ ($\chi\,,\ \xi> 90^\circ$) swings from $-90^\circ$ to $90^\circ$. A reanalysis of the same 2007 Parkes data by
\citet{2025arXiv250919446S} suggested a low-obliquity solution, with $\chi=8^\circ.4^{+5.8}_{-4.2}$ and $\xi=4^\circ.9^{+3.4}_{-2.4}$. 

\begin{table*}[ht!]
\tabletypesize{\scriptsize}
\begin{center}
\caption{RVM fit of the polarization angle. \label{tab:rvm_fit}}
\begingroup
\setlength{\tabcolsep}{14.5pt}
\renewcommand{\arraystretch}{1.6}
\begin{tabular}{c | c c c c c}
\hline\hline
\  & $\chi$ (deg) & $\xi$ (deg) & $C$ (deg) & $\gamma_0$ (deg) & $\chi^2/\mathrm{dof}$ \\
\hline
plus solution & $106.60\pm30.85$ & $157.73\pm4.04$ & $-75.35\pm1.62$ & $-69.93\pm5.15$ & $5.227/10$ \\
minus solution & $-73.40\pm30.85$ & $-22.27\pm4.04$ & $-75.35\pm1.62$ & $-69.93\pm5.15$ & $5.227/10$ \\
\hline\hline
\end{tabular}
    \endgroup
    \end{center}
\end{table*}

\section{Discussion} \label{sec:discussion}

Despite the shorter exposure, compared to those of the other magnetars targeted by \ixpe\ \cite[only 1E 1841$-$045 was observed for a comparable time;][]{rigoselli+25,stewart+25a}, we were able to perform timing, spectral, and polarization analyses of \source, obtaining significant  measurements in the $2$--$6\,\mathrm{keV}$ energy band; at higher energies, the background dominates the signal, preventing any meaningful conclusion.

The phase-averaged spectrum turns out to be well described by a single blackbody  component with temperature $\sim0.7\,\mathrm{keV}$. 
Folding the data in seven, equally-spaced phase intervals, the same spectral model provides a good representation in every phase bin. However, a variation of the temperature with the rotational phase is clearly present, which closely follows the flux modulation, while the emission radius is consistent with being a constant (see Figures \ref{fig:pp_pcube_26} and \ref{fig:param_spec_pr}). The single-peaked, quasi-sinusoidal pulse shape, the relatively large pulsed fraction, which increases with energy (see Table \ref{tab:pdpa_xspec}), and the small blackbody radius ($\sim 1.2\ \mathrm{km}$) point to a single, fairly small hot-spot. The change of $T_\mathrm{BB}$ along the rotation suggests a complex emission pattern, with radiation coming from a region on the star surface characterized by a non-uniform temperature distribution, with colder and hotter zones entering in view at different phases. 

The large phase- and energy-integrated polarization degree measured by \ixpe\ ($\approx 50\%$) places \source\ alongside the mostly polarized magnetars, 1RXS J1708 \cite[][]{zane+23} and 1E 1841$-$045 \cite[][]{rigoselli+25}, the properties of which were interpreted in terms of an atmospheric layer covering the emitting region. The same explanation possibly holds also for \source. In fact, the energy-dependent $\mathrm{PD}$ (see Figure \ref{fig:pdpa_xspec}) reaches values as high as $\approx60\%$, which is incompatible with thermal emission from a bare condensed surface around $2$--$3\,\mathrm{keV}$ and magnetospheric reprocessing via RCS at higher energies \cite[][besides, no evidence for a power-law tail was found in the \ixpe\ data]{Taverna+20}.

Although an atmosphere can explain the spectral properties and the large observed $\mathrm{PD}$ in \source, the possible presence of a local minimum in the degree of polarization between $3$ and $4\,\mathrm{keV}$, if confirmed, poses a challenge, as there is no evidence of a second spectral component or prominent spectral features that can explain such behavior 
\footnote{Especially for small emitting spots, propagation across the Quasi-Tangential (QT) region, where the photon propagation direction is nearly parallel to the magnetic field lines, can reduce the observed $\mathrm{PD}$. However, the depolarization occurs at energies well below the \ixpe\ band for magnetic fields $\gtrsim10^{14}\,\mathrm{G}$ \citep{2009MNRAS.398..515W}.}. \citet{2025arXiv250919446S} invoke mode conversion at the vacuum resonance \citep[VR; see][]{1979JETP...49..741P,1982Ap&SS..86..249K,2003MNRAS.338..233H} to explain the decrease in $\mathrm{PD}$, which they estimate to vanish at about $6\,\mathrm{keV}$, with a $90^\circ$ swing of the polarization angle at the same energy. However, our analysis shows no evidence of either a monotonic decrease with energy or a vanishing polarization at high energies. Indeed, a linear fit to the energy-dependent $\mathrm{PD}$ forcing the slope to be negative yields a $\chi^2/\mathrm{dof}=7.429/2$, and still predicts a $\mathrm{PD}\approx40\%$ at $6\,\mathrm{keV}$, while $\mathrm{PA}$ is consistent with being a constant.

\begin{figure}[t]
\includegraphics[width=0.47\textwidth]{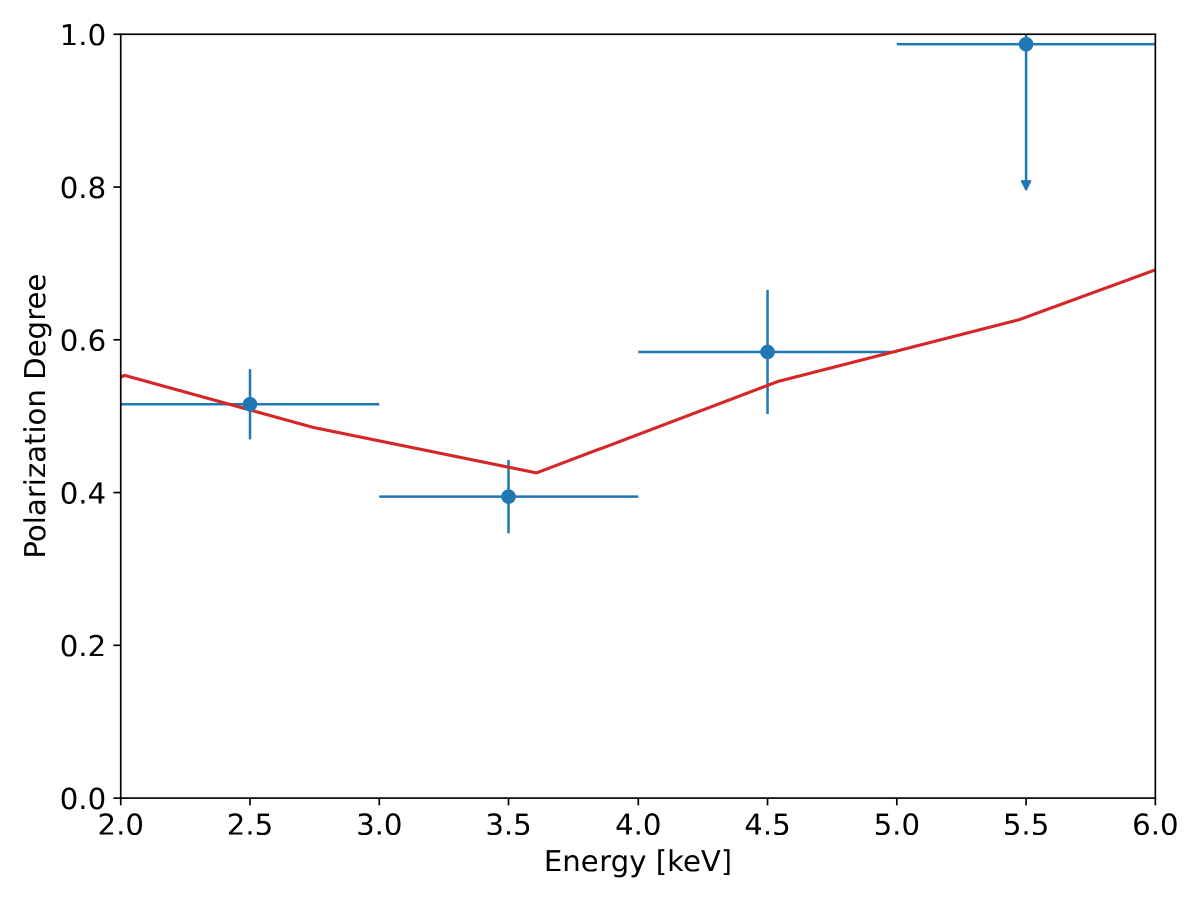}
\caption{Energy-dependent polarization degree (red) at the surface of an atmospheric patch including partial mode conversion at the VR. Here, the inclination of the magnetic field to the surface normal is $\sim 10^\circ$ and the field strength $2\times10^{14}\,\mathrm{G}$. The observed polarization degree is also shown (cyan) to facilitate visual comparison.
\label{fig:vac_res}}
\end{figure}

Still, the occurrence of mode switching at the VR does not necessarily imply a change in the dominant mode (i.e. a $90^\circ$ swing in $\mathrm{PA}$). For magnetic fields as high as $\approx2\times10^{14}\,\mathrm{G}$ (the spin-down field of \source, see \S\ref{sec:timing}), mode switching can be partial, meaning that not all O-mode photons become X-mode ones (and conversely) at the VR, with the probability of switching between modes depending on the energy of the photon, magnetic field, and density gradient in the atmosphere \cite[][]{2003MNRAS.338..233H}. This results in a local minimum in $\mathrm{PD}$, while the polarization angle remains constant with energy \citep[][and references therein]{2024MNRAS.528.3927K,2025ApJ...987..113K}. To test whether (partial) mode conversion at the vacuum resonance can produce a behavior of $\mathrm{PD}$ like that suggested by \ixpe\ between $3$ and $4\,\mathrm{keV}$, we performed some simulations following the method detailed in \citet{2024MNRAS.528.3927K}. Figure \ref{fig:vac_res} shows the energy-dependent behavior of the polarization degree in the case of emission from  a single atmospheric patch, assuming a field of strength $B=2\times10^{14}\,\mathrm{G}$, inclined by $\sim 10^\circ$ wrt the surface normal and a probability threshold of $P_\mathrm{th}=0.9$ \footnote{The magnetic field used in the simulation of Figure \ref{fig:vac_res} is somewhat smaller than the polar value inferred in \source, because convergence issues with our atmospheric code prevented us from reaching higher values of $B$. We successfully produced a run with an aligned field of $B=4\times10^{14}\,\mathrm{G}$ on a coarser energy grid which, albeit noisier, is consistent with the results shown in Figure \ref{fig:vac_res}.}
Although a more detailed model is required before any definitive conclusion can be drawn, the results reported in Figure \ref{fig:vac_res} suggest that partial mode conversion at the VR can produce dips in $\mathrm{PD}$ at the energies observed by \ixpe\ in \source.

The phase variation of $\mathrm{PD}$ is statistically consistent with a constant within $1\sigma$ uncertainties. However, the most probable values hint to an anti-correlation of $\mathrm{PD}$ with the pulse profile in the total band ($2$--$6\, \mathrm{keV}$), and also in the $2$--$3$ and $3$--$6\,\mathrm{keV}$ energy intervals (see the top panels of Figures \ref{fig:pdpa_xspec_pr} and $\ref{fig:pdpa_xspec_pr_2336}$). On the other hand, the polarization angle traces a clean sinusoidal modulation with the rotational phase, which can be well fitted with a rotating vector model (see Figure \ref{fig:rvm_sim_fit}). This pattern 
is common to other magnetars observed by \ixpe\ and has been interpreted as a signature of vacuum birefringence in the external region close to the star \cite[][]{taverna+22,zane+23,heyl+24}. While $\mathrm{PD}$ bears the imprint of the (complex) thermal and magnetic maps on the star surface, much as the pulse profile, $\mathrm{PA}$ is determined at larger distances where the topology of the field is largely dipolar. This is a consequence of vacuum birefringence, which forces the polarization vectors to follow the magnetic field direction up to the polarization-limiting radius \cite[$r_\mathrm{pl}\approx 100\,R_\mathrm{NS}$ for magnetars;][]{2003MNRAS.342..134H,2015MNRAS.454.3254T}.

According to \citet{2025arXiv250919446S}, who fitted the polarization angle measured in the radio band with the Parkes/Murriyang telescope, the star is an almost aligned rotator, seen nearly along the magnetic axis. In this case, for photons coming from regions very close to the pole, the polarization vectors at emission would co-rotate with the star, following the magnetic field near the pole. As a consequence, if vacuum birefringence is neglected, the phase-averaged polarization degree observed at infinity should be significantly lower than what is expected when a proper account for QED effects is made, even assuming that emitted photons are polarized mostly in one mode. This is one of the points that led \citet{2025arXiv250919446S} to claim that the high phase-averaged and phase-resolved polarization degree detected by \ixpe\ in \source\ is a compelling test of vacuum birefringence.

\begin{figure}[t]
\includegraphics[width=0.47\textwidth]{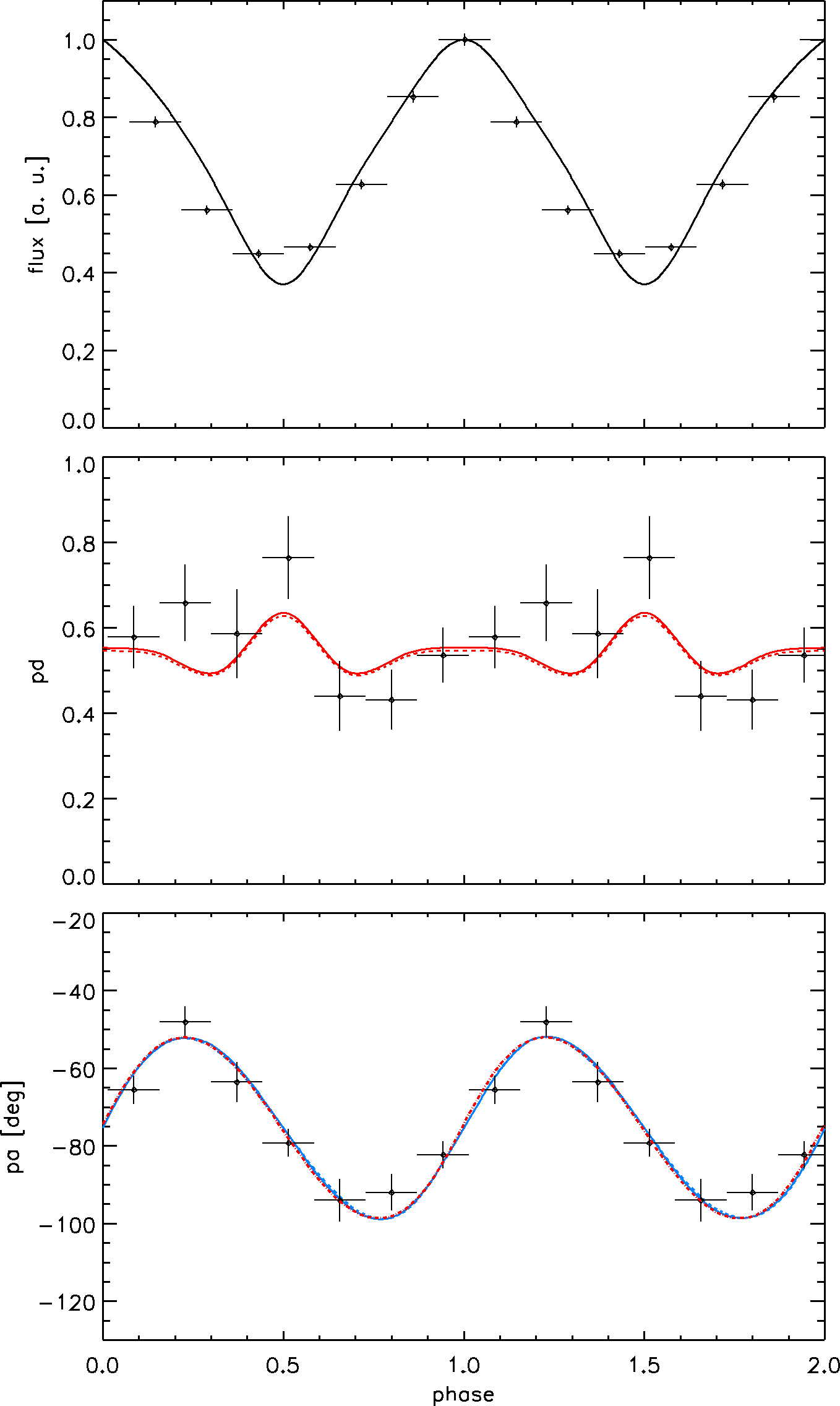}
\caption{Pulse profile (normalized to the maximum, top), phase-dependent $\mathrm{PD}$ (middle) and $\mathrm{PA}$ (bottom) as obtained from the ray-tracing code discussed in \citet{2015MNRAS.454.3254T}, using $\chi_-$ and $\xi_-$ 
(red dash-dotted line in the bottom panel). The emitting region is a circular spot at the magnetic pole (see text for details). Solid and dashed lines show the results obtained with and without including vacuum birefringence effects, respectively. \ixpe\,data are also shown (black points with error bars). Both flux and $\mathrm{PD}$/$\mathrm{PA}$ points have been phase-shifted back by $\approx0.65$ cycles wrt Figures \ref{fig:pp_pcube_26}, \ref{fig:pdpa_xspec_pr} and \ref{fig:pdpa_xspec_pr_2336} to facilitate visual comparison.
\label{fig:raytracer_sim}}
\end{figure}
Our RVM fit of $\mathrm{PA}$ measured by \ixpe\ produces a different geometry, with a net inclination between the magnetic and rotation axes of $\sim 20^\circ$ and the LOS tilted by $\sim 75^\circ$ wrt the rotation axis. The nearly aligned geometries proposed by \citet{2025arXiv250919446S} are,
according to our posteriors, excluded with significance $\gtrsim 90\%$.

In the geometry we found, the difference in the degree of polarization obtained with or without accounting for vacuum birefringence is expected to be negligible 
because of the limited extent of the emission region and of the strong magnetic field of the source. In such cases, in fact, the change in the magnetic field direction across the emitting area is quite small, mimicking the conditions that would generally occur at a large distance from the source \cite[$r_\mathrm{pl}\approx100R_\mathrm{NS}$,][]{2009MNRAS.398..515W}.

To provide a quantitative estimate, we performed a simulation using the ray-tracing code discussed in \citet{2015MNRAS.454.3254T}. General relativistic corrections, both in the photon trajectories and in the magnetic field, are accounted for. We assumed a neutron star mass and radius of $M=1.4\,M_\odot$ and $R=12\ \mathrm{km}$, respectively, and a (polar) value of the dipole field $B\sim5\times10^{14}\,\mathrm{G}$. The emission comes from a circular spot centered on the magnetic pole with semi-aperture $\theta\approx6^\circ$, corresponding to $R_\mathrm{BB}\approx1.2\,\mathrm{km}$ (as evaluated by a distant observer, see Table \ref{tab:spec_xspec}). The spot is divided into a hot ($kT_\mathrm{h}=0.87\,\mathrm{keV}$, $\theta_\mathrm{h}\approx 3^\circ$) cap with a large intrinsic polarization ($80\%$), surrounded by a warm ($kT_\mathrm{w}=0.76\,\mathrm{keV}$) annulus, extending from $3^\circ$ to $6^\circ$, with a lower polarization ($30\%$); the rest of the surface is cold, at $kT_\mathrm{c}=0.15\,\mathrm{keV}$. Both regions emit isotropic blackbody radiation and the (constant) polarization degrees were chosen in such a way to recover the observed $\mathrm{PD}$ in the \ixpe\ band. All previous quantities are measured by a stationary observer at the star surface, unless explicitly stated otherwise.

The results, shown in Figure \ref{fig:raytracer_sim} for our geometry, are qualitatively in agreement with the \ixpe\ measurements in the $2$--$6\,\mathrm{keV}$ band, and 
successfully reproduce the single-peaked pulse profile, a nearly constant phase-dependent polarization degree (weakly anti-correlated with the light curve), and a clear sinusoidal oscillation of the polarization angle. 
Actually, emission from a more realistic magnetized atmosphere model is not expected to be isotropic. In order to test if and to which extent magnetic beaming can influence the results presented in Figure \ref{fig:raytracer_sim}, we computed some models replacing the Planck function, $B_\nu$, with  $B_\nu f(\mu)$, where $\mu = \cos\theta_\mathrm{Bk}$ is the cosine of the angle between the local magnetic field and the ray. The beaming factor $f(\mu)$ was taken to qualitatively reproduce the angular pattern of the intensity emerging from a strongly magnetized atmosphere \citep[e.g.][]{2001ApJ...563..276O}. However, only small deviations in the pulse profiles were found: fractional differences in the pulsed fraction are typically $\lesssim 10\%$) while both $\mathrm{PD}$ and $\mathrm{PA}$ are pretty much unchanged.
At any rate, and as for the model shown in Figure \ref{fig:vac_res}, we stress that
this simple model is not intended to faithfully match all the observed features, but rather to demonstrate that the derived viewing geometry and emission configuration produce results consistent with the observations.
As expected, the polarization degree calculated with or without considering vacuum birefringence (solid and dashed lines in the central panel) is nearly indistinguishable, so no compelling argument about the presence of vacuum birefringence can be made. 

In this respect, it is worth drawing some comparisons between our analysis and that of the magnetar 1RXS J1708 presented in \citet{zane+23}. Here we have shown that a simple emission model, which relies on the source geometry inferred from the RVM fit of the \ixpe\ polarization angle, can broadly reproduce the main X-ray properties of the source. In the case of 1RXS J1708, a somewhat reverse approach was followed. First, possible emission geometries capable of explaining the (energy-dependent) spectro-polarimetric data were explored, and then these were contrasted with the phase variation of  
$\mathrm{PA}$ and with the RVM best fit. The agreement they found is only qualitative, not surprisingly given that no real fit of $\mathrm{PA}$ with their models was attempted. 

The relatively large half-opening angle of the radio beam required by our X-ray-favored geometry to match the observed duty cycle ($\rho\sim 22^\circ$) does not necessarily invalidate the model. Although the empirical relation $\rho\sim 5^\circ/\sqrt{P}$ predicts a much narrower beam ($\sim 3.4^\circ$ for a period $2.1\, \mathrm{s}$), this scaling is derived primarily from ordinary rotation-powered pulsars. \source\ is a high-energy magnetar with a rotational loss rate ($\dot E\sim 9\times 10^{34}\, \mathrm{erg/s}$) and magnetic field strength far exceeding the population used to establish such relations. Given the significant scatter in beam widths observed even among ordinary pulsars \citep[see e.g.][]{2019MNRAS.485..640J}, and the unique magnetospheric physics of magnetars, an opening angle $\rho\sim 22^\circ$ cannot be ruled out.

\section{Conclusions}\label{sec:concl}
In this work, we presented an analysis of the $500\,\mathrm{ks}$ \ixpe\ observation of the AXP \source. 
We found that the source X-ray emission is essentially thermal, with only one blackbody component contributing at each phase of the star rotation, and comes from a fairly limited region on the star surface ($R_\mathrm{BB}\approx 1.2\, \mathrm{km}$) with non-uniform properties, as suggested by the modulation of the temperature with phase. 

Both the relatively high phase-averaged and phase-resolved polarization degrees are consistent with emission from a magnetized atmospheric layer covering (part of) the star surface. The constancy of the polarization angle with energy implies that radiation is polarized in the same mode throughout the \ixpe\ band, as predicted for an ultra-magnetized atmosphere, the emission of which is expected to be X-mode dominated \cite[][]{2024MNRAS.528.3927K}. The only potential evidence of QED effects in the atmospheric medium (vacuum+plasma) comes from the dip in the degree of polarization that is possibly present around $3$--$4\,\mathrm{keV}$, which may indeed be a signature of partial mode conversion at the vacuum resonance.

On the other hand, the rotating vector model provides a very good interpretation of the modulation in phase of the polarization angle measured by \ixpe\ and this hints at the presence of vacuum birefringence in the star magnetosphere \citep[][]{taverna+22}. In the geometry we inferred from the RVM fit, \source\ is an inclined rotator seen nearly perpendicular to the spin axis and this is consistent with the observed pulse profile, assuming isotropic thermal emission from a polar hot spot on the star surface. The geometry derived by \citet{2025arXiv250919446S} from radio data points to an almost aligned rotator seen close to the pole, which is unlikely on the basis of \ixpe\ data alone. 
Still, given that it falls at the boundary of our $3\sigma$ contour (see Figure \ref{fig:post}), it cannot be excluded, potentially removing the tension between the measures in the two bands. On the other hand, since the rotating vector model constrains the source geometry independently of the assumed emission model, the magnetic field topology in the regions responsible for the radio and the X-ray emission may be different. Further multi-wavelength observations of \source\ are needed to clarify this issue.

The actual source geometry has an impact on the conclusions one can draw about the presence of QED effects. In the case of an inclined geometry, as the one derived here, the polarization degree of radiation emitted from a small spot is maintained up to the observer for magnetar-like fields, and no claim about the presence of vacuum birefringence can be made, no matter how large the polarization degree is. In this respect, \source\ is not different from other persistent magnetars observed in polarized X-rays (e.g., 1RXS J1708 and 1E 1841$-$045), confirming that such sources are not ideal targets for (indirect) tests of QED, unless the emission comes from a reasonably extended portion of the surface. 

A significant step forward is expected from observations of transient magnetars in outburst, during which emission is indeed expected to emerge from wide hot spots. Another promising class of sources is that of X-ray Dim Isolated Neutron Stars (XDINSs), which are also strongly magnetized \cite[albeit with fields typically an order of magnitude lower than those of magnetars, see][]{2009ASSL..357..141T}, the prototypical and brightest member of which, RX J1856.5$-$3754, exhibits a very low X-ray pulsed fraction \cite[][]{2012A&A...541A..66S}, making them key targets for testing vacuum birefringence with future instruments capable of probing polarization at sub-keV energies  \cite[see e.g.][for predictions in the soft X-rays and \citealt{2017MNRAS.465..492M} for a potential QED detection in the optical]{2025ApJ...987..113K}. Next-generation X-ray polarimeters, like {\it eXTP} \cite[the Enhanced X-ray Timing and Polarimetry mission,][]{2025SCPMA..6819502Z}, {\it EXPO} (the Enhanced X-ray Polarimetry Observatory, recently proposed for the ESA medium-size mission call), with their superior sensitivity and fast-repointing capabilities in the medium-to-high energy X-ray band, and GOSoX \cite[the Globe Orbiting Soft X-ray polarimeter concept,][]{2021SPIE11444E..2YM}, which will observe in the soft-X band, will hopefully provide the definitive answer on the long-sought-after test of vacuum birefringence, finally bridging the gap between fundamental physics and astrophysical observations of neutron stars.

\begin{acknowledgments}
We gratefully acknowledge an anonymous referee for their constructive criticism and helpful remarks. This research was supported by the International Space Science Institute (ISSI) in Bern, through the International Team project 25-657 ``Polarimetric Insights into Extreme Magnetism''.
The work of R. Ta., R. Tu. and L. M. is partially funded by the PRIN 2022 - project 2022LWPEXW - ``An X-ray view of compact objects in polarized light'', European Union funding - Next Generation EU, Mission 4 Component 1, CUP C53D23001180006.
R.M.E. K. is supported by The Science and Technology Facilities Council (STFC) via a PhD studentship (grant number ST/W507891/1). GL. I. and S. M. acknowledge financial support from INAF through the Bando Ricerca Fondamentale INAF 2024, Large Grant ``TULiP''   
and GO Grant ``Toward Neutron Stars Unification'', respectively. A. B. acknowledges support through the European Space Agency (ESA) research fellowship program. 
\end{acknowledgments}




\bibliography{sample701}{}
\bibliographystyle{aasjournalv7}



\end{document}